\documentclass[
preprint,
superscriptaddress,
groupedaddress,
 amsmath,amssymb,
onecolumn,
showkeys
]{revtex4-2}
\usepackage{hyperref}
\usepackage{graphicx}
\usepackage{dcolumn}
\usepackage{xcolor}
\usepackage{bm}
\usepackage[title]{appendix}
\usepackage{hyperref}

\begin{document}
\title{Hamiltonian formulation of X-point collapse in an extended magnetohydrodynamics framework}

\author{Hamdi M. Abdelhamid}
\email{hamdi_mktprg@mans.edu.eg}
\affiliation{Physics Department, Faculty of Science, Mansoura University, Mansoura, 35516, Egypt}
\affiliation{Physics Department, Faculty of Science, New Mansoura University, New Mansoura City, 35742, Egypt}

\author{Manasvi Lingam}
\email{mlingam@fit.edu}
\affiliation{Department of Aerospace, Physics and Space Sciences, Florida Institute of Technology, Melbourne, FL 32901, USA}

\begin{abstract}
The study of X-point collapse in magnetic reconnection has witnessed extensive research in the context of space and laboratory plasmas. In this paper, a recently derived mathematical formulation of X-point collapse applicable in the regime of extended magnetohydrodynamics (XMHD) is shown to possess a noncanonical Hamiltonian structure composed of five dynamical variables inherited from its parent model. The Hamiltonian and the noncanonical Poisson bracket are both derived, and the latter is shown to obey the requisite properties of antisymmetry and the Jacobi identity (an explicit proof of the latter is provided). In addition, the governing equations for the Casimir invariants are presented, and one such solution is furnished. The above features can be harnessed and expanded in future work, such as developing structure-preserving integrators for this dynamical system.
\end{abstract}
\maketitle

\section{Introduction}\label{SecIntro}

Magnetohydrodynamics (MHD) constitutes the bedrock of fluid descriptions of plasmas, owing to its powerful combination of mathematical simplicity and broad physical applicability \cite{TC62,KT73,TS97,BS03,PMB08,JPF08,JPF14,SG16,GB17,GKP19}. MHD has been extensively deployed to study the physical dynamics of plasmas ranging from traditional areas such as nuclear fusion \cite{KT73,JPF08,JPF14,HZ14,LZ19}, astrophysics \cite{TS97,MG03,RK05,BS12,CV15,PAD17}, space and solar physics \cite{KR95,MK01,MJA04,PF04,SM11,EP14,RLS16,GB17,BT22} to emerging and rapidly evolving new disciplines like (exo)planetary science and astrobiology \cite{DJL18,LL19,GAB20,ML21}.

However, it must be appreciated that MHD does not always offer a realistic description of plasma dynamics, since this model has been derived under a specific set of approximations \cite{GKP19}. For instance, MHD typically diminishes in accuracy when addressing high-frequency phenomena and small-scale structures in highly conductive plasmas. In such instances, it is necessary to proceed ``beyond MHD'' and employ models with two-fluid effects such as the Hall current and electron inertia. These models are often grouped together under the umbrella of extended magnetohydrodynamics (XMHD) \cite{RL59,JG73,KM14,KLM14,JWB17,ET17,GKP19,CB24}, of which Hall magnetohydrodynamics (HMHD) is the simplest \cite{Light60,RTC79,TT83,EW86,Tur86,Holm87}.

A fundamental phenomenon in plasma physics is magnetic reconnection, which entails changes in magnetic field line topology (due to their breaking and reconnection) accompanied by energy conversion \cite{BKS85,DB94,DB00,AB04,ZY09,YKJ10,JDJ22,MY22}. This process has been invoked to explain myriad phenomena spanning solar flares \cite{HSH11,SM11} and coronal mass ejections \cite{PC11,WH12} to magnetospheres \citep{KS10,ZA13,HC20,TPF21}, high-energy astrophysical processes \citep{CV15,CA21,JDJ22,GLZ24,LHG24}, and nuclear fusion \cite{YKJ10,MY22}. While certain resistive MHD reconnection models, such as those mediated by plasmoids \cite{TS97,BHY09,CLHB,CLH17,DWH22} or turbulence \citep{LL99,LVK12,LEV15,LEJ20}, are held to be sufficiently fast, the synthesis of laboratory experiments, astronomical observations, and numerical models has established that XMHD (especially HMHD) facilitates fast reconnection in a plethora of environments \cite{BSD97,SDR01,PBC02,AB04,Fitz04,JDH05,SDS07,SC08,DSS08,ZY09,BKS11,CB16,MY22,JYF23,LHG24}.

An important component of magnetic reconnection involves studying the dynamics of X-point collapse (at which locations the magnetic field is zero), which facilitates the formation of current sheets \cite{JWD53,SHB88,KS10,YKJ10,EP14,LMS15,PP22}. The phenomenon of X-point collapse and current sheet formation has been thoroughly investigated in both analytical and numerical publications using both MHD \cite{IS67,SIS71,BO84,CW92,TP93,PTV94,LCD05,EP14,PP22}, and collisionless models with XMHD effects \cite{BPS92,SC06,TH07,CL08,BKS09,PLP08,BKS11,TPD12,VT14}. In this paper, we will explore the nonlinear dynamical system elucidated in Refs. \cite{YEL07,LM15a,LM15b,AZJ18,AJB19,AJZ19}, which captures X-point collapse in XMHD. In particular, we will focus on uncovering the rich noncanonical Hamiltonian structure of this theoretical model. 

Our emphasis on the Hamiltonian structure goes beyond mathematical elegance. The advantages associated with (noncanonical) Hamiltonian formulations are manifold \cite{PJM82,JEM92,PJM98,PJM05,PJM09,HSS09,ML15,SM16}, some of which are adumbrated as follows: (1) they can help unearth hidden connections between outwardly disparate models, which occurred in the case of XMHD \cite{AKY15,LMM15}; (2) they enable the extraction of topological invariants of the dynamical system \cite{HMR85,JEM92,PJM98}, as shown for XMHD \cite{AKY15,LMM16,MLL17}; (3) they enable the rigorous determination of both linear and nonlinear stability of equilibria \cite{PJM82,HMR85,AMP13,KTM20,ET22}; (4) they enable the systematic derivation of ``offspring'' models from the ``parent'' that retain the requisite symmetries and conservation laws, as established for XMHD to pick a single example \cite{KLM14,DAML16}; and (5) they can be suitably discretized, thereby enabling the first-principles derivation of symmetry-preserving algorithms with guaranteed high accuracy in a variety of fields \cite{CS90,JMS92,LS94,LR01,FQ10,OO22}, including \cite{MK13,ZQB14,CEE16,PJM17}. This last aspect, in particular, can be valuable for future work on X-point collapse, which would be characterized by (near-)singularities in simulations.

The outline of the paper is as follows. In Section \ref{SecXMHD}, we review the formulation of reduced XMHD, which is essential for understanding the model of Refs. \cite{AZJ18,AJZ19} utilized herein, building on the previous work by Refs. \cite{YEL07,LM15a,LM15b}. Subsequently, we present the nonlinear ODEs for X-point collapse from Ref. \cite{AZJ18} in Section \ref{SecJanda}, and then elucidate their corresponding noncanonical Hamiltonian structure in Section \ref{SecHamilton} and Appendix \ref{apendixA}. At the end, we briefly summarize our findings in Section \ref{SecConc}.

\section{Reduced extended MHD: the Hamiltonian formulation}\label{SecXMHD}
In the incompressible limit, reduced extended magnetohydrodynamics (RXMHD) can be obtained by casting the remaining independent variables -- namely, the velocity and the magnetic field $\left(\mathbf{V},\mathbf{B}\right)$ --  into a Clebsch-potential parametrization \cite{JS59,SW68,Yosh09}. The detailed calculation is described in Ref. \cite{Grasso2017}, owing to which we restrict ourselves to a synopsis, and we adopt the same notation and approach as the aforementioned reference. Before proceeding, we remark that Hamiltonian models that are either descended from or partially related to RXMHD have been used to study magnetic reconnection \citep{CGP98,KLP98,GPP99,GCPP,PBC02,TMW08,TMG10}, intertwining the two themes introduced in Section \ref{SecIntro}.

The magnetic field, which must obey the divergence-free condition of $\nabla \cdot \mathbf{B} = 0$, is cast in two dimensions into the following form:
\begin{equation}
\label{var1}
\mathbf{B}\left(x,y,t\right)=\nabla\psi\left(x,y,t\right)\times\hat{z}+b\left(x,y,t\right)\hat{z},
\end{equation}
whereas an exactly similar decomposition is viable for the 2D velocity $\mathbf{V}$ in the limit of incompressibility (i.e., $\nabla \cdot \mathbf{V} = 0$),
\begin{equation}
\label{var2}
\mathbf{V}\left(x,y,t\right)=-\nabla\phi\left(x,y,t\right)\times\hat{z}+v\left(x,y,t\right)\hat{z}.
\end{equation}
where $\psi$ and $\phi$ denote the so-called flux and stream functions; we comment further on the mathematical form of this equation in Section \ref{SecJanda}. Moreover, on inspection, it is apparent that $b$ and $v$ are $z$-components of the magnetic and velocity fields, respectively. Under this representation, we can define the current density $\mathbf{J}$ as
\begin{equation}
\label{Current}
\mathbf{J}=\nabla\times\mathbf{B}=\nabla b\times\hat{z}-\nabla^{2}\psi\hat{z},
\end{equation}
In a suitably normalized format, the momentum equation and the generalized Ohm's low for the full XMHD model, in the absence of the myriad of (nonlinear) dissipative effects \cite{LHP17}, are given by \cite{KM14,KLM14,GKP19}:
\begin{eqnarray}
\label{IMom}
\frac{\partial\mathbf{V}}{\partial t}  + \left(\mathbf{V}.\nabla\right)\mathbf{V} = \mathbf{J}\times\mathbf{B}-d^{2}_{e}\left(\mathbf{J}\cdot\nabla\right)\mathbf{J},
\end{eqnarray}
\begin{eqnarray}
\label{IOhm}
\mathbf{E}+\mathbf{V}\times\mathbf{B}=d_{i}\mathbf{J}\times\mathbf{B}&+&d^{2}_{e}\left[ \frac{\partial \mathbf{J}}{\partial t}+(\mathbf{V}\cdot\nabla)\mathbf{J}+\left(\mathbf{J}\cdot\nabla\right)\mathbf{V}\right] - d_{i}d^{2}_{e}\left(\mathbf{J}\cdot\nabla\right)\mathbf{J},
\end{eqnarray}
where we have dropped the total pressure in (\ref{IMom}) and the electron pressure in (\ref{IOhm}), respectively, because these pure gradient terms vanish when the curl is taken, and they do not contribute explicitly to the dynamics of the simple model delineated hereafter. The quantities $d_i$ and $d_e$ refer to the (normalized) ion and electron skin depths, respectively. The $\hat{z}$ component of \eqref{IMom}, after some algebraic manipulation, yields the dynamical equation:
\begin{eqnarray}
\label{R1}
\frac{\partial v}{\partial t}&=&-\left[\phi,v\right]+\left[b,\psi\right]-d^{2}_{e}\left[b,\nabla^{2}\psi\right],
\end{eqnarray}
where we have introduced the standard Poisson bracket:
\begin{eqnarray}
\left[f,g\right]=\nabla f\times\nabla g \cdot \hat{z},
\end{eqnarray}
encountered often in 2D plasma fluid models. In a similar fashion, the $\hat{z}$ component of the curl ($\nabla \times$) of \eqref{IMom} gives
\begin{eqnarray}\label{R2}
\frac{\partial \left(\nabla^{2}\phi\right)}{\partial t}&=&-\left[\phi,\nabla^{2}\phi\right]-\left[\nabla^{2}\psi,\psi\right]-d_e^{2}\left[b,\nabla^{2}b\right].
\end{eqnarray}
Likewise, we can write the $\hat{z}$ component of \eqref{IOhm} after much simplification as
\begin{eqnarray}
\label{R3}
-\frac{\partial \psi}{\partial t}+\left[\psi,\phi\right]&=&d_{i}\left[b,\psi\right]-d^{2}_{e}\frac{\partial }{\partial t}\nabla^{2}\psi+d^{2}_{e}\left[\nabla^{2}\psi,\phi\right] + d^{2}_{e}\left[v,b\right]-d_{i}d^{2}_{e}\left[b,\nabla^{2}\psi\right],
\end{eqnarray}
where we had to use the relation, $E_{z} = -\partial \psi/\partial t$, which follows from harnessing the definition of the electric field. Finally, on taking the curl ($\nabla \times$) of \eqref{IOhm}, the $\hat{z}$ component of the ensuing vectorial equation simplifies to
\begin{eqnarray}
\label{R4}
-\frac{\partial b}{\partial t}&&+\left[v,\psi\right]-\left[\phi,b\right]=d_{i}\left[\psi,\nabla^{2}\psi\right]-d^{2}_{e}\frac{\partial }{\partial t}\nabla^{2}b + d^{2}_{e}\left[\nabla^{2}\phi,b\right] \nonumber \\
&& \hspace{1.3in} +d^{2}_{e}\left[\nabla^{2}b,\phi\right]-d_{i}d^{2}_{e}\left[b,\nabla^{2}b\right].
\end{eqnarray}

For the system of equations given by \eqref{R1}, \eqref{R2}, \eqref{R3}, and \eqref{R4}, the Hamiltonian (total energy in this case) is found to be
\begin{equation}
\label{RE}
H:= \frac{1}{2} \int d^{2} x \left(- \phi \omega - \psi^{\ast} \nabla^2 \psi + b^{\ast} b + v^{2}\right),
\end{equation}
where we have introduced the following auxiliary variables for the sake of simplicity:
\begin{equation}\label{omdef}
 \omega=\nabla^{2}\phi, 
\end{equation}
\begin{equation}
\psi^{\ast}=\psi-d^{2}_{e}\nabla^{2}\psi,
\end{equation}
\begin{equation}
b^{\ast}=b-d^{2}_{e}\nabla^{2}b.
\end{equation}
In \eqref{RE}, the first and last terms mostly embody the ion kinetic energy, whereas the remaining two terms capture the residual electron kinetic energy and the magnetic energy \cite{KLM14}.

It was shown in Ref. \cite{Grasso2017} that, in addition to the Hamiltonian \eqref{RE} formulated here, the system of equations derived for RXMHD possesses a noncanonical Hamiltonian formulation, which we shall not repeat here. 

\section{Self-similar reduction of RXMHD}\label{SecJanda}

By starting with RXMHD, it is possible to extract a system of nonlinear ODEs (as opposed to PDEs) that captures the necessary dynamics for X-point collapse. This procedure was worked out in Ref. \cite{AZJ18}, owing to which we only sketch the steps -- by mirroring the derivation and notation to facilitate easier reading -- and refer the reader to that publication for a detailed treatment. In particular, we underscore that Ref. \cite{AZJ18} sought to motivate the physical rationale behind the choice of (\ref{psi})--(\ref{b}), owing to which the forenamed paper may be consulted for an in-depth description of the same, in tandem with Ref. \cite{PKR95} who delineated a simpler model of X-point dynamics under self-similar assumptions.

A planar X-point configuration was assigned for the magnetic field (see Ref. \cite{MSU63}), thereby leading to the ansatz of
\begin{equation}\label{psi}
\psi\left(t,x,y\right)=\alpha_{1}\left(t\right) x^{2}-\alpha_{2}\left(t\right) y^{2}+ 2 d^{2}_{e}\left(\alpha_{1}\left(t\right)-\alpha_{2}\left(t\right)\right) +2\eta \int\left(\alpha_{1}\left(t\right)-\alpha_{2}\left(t\right)\right) dt,
\end{equation}
where $\eta$ is the plasma resistivity, a dissipative effect, and is therefore equal to zero in the Hamiltonian limit. Next, the assumption of a stagnation-point flow was employed to yield the following relation:
\begin{equation}\label{phi}
\phi\left(t,x,y\right)= -\gamma\left(t\right) x\, y, 
\end{equation}
where we further comment on $\gamma$ at the end of this section. On inputting this expression into (\ref{var2}), we find that the $x$- and $y$-components of the velocity are respectively proportional to $x$ and $y$, verifying the mathematical description of a stagnation point \citep[Chapter 2.3]{DR06}. The $z$-component of the velocity was determined to obey the ansatz:
\begin{equation}\label{v}
v\left(t,x,y\right)=\beta_{1}\left(t\right) x^{2}+\beta_{2}\left(t\right) y^{2}+ +2 \nu \int\left(\beta_{1}\left(t\right)+\beta_{2}\left(t\right)\right) dt,
\end{equation}
where $\nu$ is the plasma viscosity, another dissipative effect, that would vanish when the Hamiltonian limit is computed. The prescription (\ref{v}) fulfills (\ref{R1}), after using the expressions for the other variables. Finally, the $z$-component of the magnetic field was chosen to exhibit the quadrupolar structure found in Hall reconnection \cite{WBM00}, thereby amounting to
\begin{equation}\label{b}
b\left(t,x,y\right)=b\left(t\right) x y.
\end{equation}

On plugging the preceding ansatzen into the system \eqref{R1}-\eqref{R4}, the latter set of equations can be reduced to the following system of ordinary differential equations: 
\begin{equation}\label{alpha1}
\frac{d }{d t}\alpha_{1}\left(t\right)=2\big(d_{i}b\left(t\right)-\gamma\left(t\right)\big)\alpha_{1}\left(t\right)-2 d^{2}_{e} \beta_{1}\left(t\right) b\left(t\right) .
\end{equation}
\begin{equation}\label{alpha2}
\frac{d }{d t}\alpha_{2}\left(t\right)=2\big(\gamma\left(t\right)-d_{i} b\left(t\right)\big)\alpha_{2}\left(t\right)-2 d^{2}_{e} \beta_{2}\left(t\right) b\left(t\right),
\end{equation}
\begin{equation}\label{beta1}
\frac{d }{d t}\beta_{1}\left(t\right)=-2\gamma\left(t\right) \beta_{1}\left(t\right)-2 \alpha_{1}\left(t\right) b\left(t\right),
\end{equation}
\begin{equation}\label{beta2}
\frac{d }{d t}\beta_{2}\left(t\right)=2\gamma\left(t\right) \beta_{2}\left(t\right)-2 \alpha_{2}\left(t\right) b\left(t\right),
\end{equation}
\begin{equation}\label{b2}
\frac{d }{d t}b\left(t\right)=-4\alpha_{1}\left(t\right) \beta_{2}\left(t\right)-4\alpha_{2}\left(t\right) \beta_{1}\left(t\right).
\end{equation}

Before moving on the Hamiltonian formulation of (\ref{alpha1})--(\ref{b2}), a couple of important comments are worth noting.
\begin{itemize}
    \item In Section \ref{SecXMHD}, we presented the nondissipative limit of RXMHD. It turns out, however, that even if (\ref{psi})--(\ref{b}) are plugged into the \emph{dissipative} version of RXMHD, we would still end up with (\ref{alpha1})--(\ref{b2}), as shown in Ref. \cite{AZJ18}.
    \item The first terms on the RHS of (\ref{alpha1}), (\ref{alpha2}), (\ref{beta1}), and (\ref{beta2}) are slightly different their counterparts of Ref. \cite{AZJ18}. The reason is that we have adopted the standard Clebsch-potential parametrization in (\ref{var2}), whereas Ref. \cite{AZJ18} does not have the negative sign, as illustrated below:
    \begin{equation}
    \label{var2v2}
    \mathbf{V}\left(x,y,t\right)=\nabla\phi\left(x,y,t\right)\times\hat{z}+v\left(x,y,t\right)\hat{z}.
    \end{equation}
    In other words, the difference in the final dynamical equations reported above arises from merely a difference in the choice of sign convention.
    \item Although we have five dynamical equations, describing the temporal evolution of the five dynamical variables of interest, there are actually six time-dependent variables in \eqref{alpha1}-\eqref{b2}. The ``odd'' one in this group is $\gamma$, which is argued to be a freely specifiable function by Ref. \cite{AZJ18}. We briefly revisit this point below, as well as in Section \ref{SSecCasimir}, but for now, we clarify that we treat the time-dependent variable $\gamma(t)$ as an arbitrary function of the four dynamical variables, namely, $\gamma(t) \equiv \gamma\left(\alpha_1,\alpha_2,\beta_1,\beta_2,b,t\right)$, which includes the key possibility that it can be independent of all of them.
    \item An issue with self-similar solutions of the kind described by (\ref{psi})--(\ref{b}) is that, when integrated over the entire spatial domain, they would correspond to a formal case of infinite energy (e.g., Ref. \cite{BO84}); in other words, achieving a finite energy would require the truncation of the domain to a finite region, which introduces subtleties regarding flux terms (as sketched below). The reason for the X-point collapse may be attributable to possibly an influx of energy from infinity.
    \item If we calculate $\omega$, given by (\ref{omdef}), via applying (\ref{phi}), it is found that this quantity would vanish for this particular ansatz. Hence, the first term on the RHS of (\ref{RE}) as it stands would yield a kinetic energy (density) component equal to zero, which seems unrealistic. In contrast, if we perform an integration by parts of this term, we would arrive at $|\nabla \phi|^2$ plus an additional flux term that must be evaluated at the boundary (if the region is finite). As with the preceding point, it can be shown that $|\nabla \phi|^2$ would be rendered infinite over an infinite domain, and could thus perhaps warrant a choice of $\gamma = 0$. The claim that $\gamma$ appears to be undetermined by Ref. \cite{AZJ18} may be a consequence of the ambiguity surrounding (\ref{RE}) for the aforementioned ansatz (and domain), and might indicate that the self-similar ansatz does not yield a unique solution of the initial set of governing equations, conceivably on account of the elision or suppression of proper boundary conditions.
\end{itemize}

\section{Hamiltonian Formulation}\label{SecHamilton}

We are now in a position to elucidate the noncanonical Hamiltonian structure of the above ordinary differential equations \eqref{alpha1}-\eqref{b2}. Prior to this endeavor, we point out a simpler system for X-point dynamics derived under the assumption of self-similarity was shown to possess a Hamiltonian structure by Ref. \cite{PKR95}.

The reader is referred to Refs. \cite{PJM82,PJM98,SM16} for a review of noncanonical Hamiltonian dynamics. In a nutshell, suppose that the set of $N$ dynamical variables are denoted by $u_i$, with $i=0,1,2,\dots N-1$; in our system \eqref{alpha1}-\eqref{b2}, note that $N=5$ (i.e., there are $5$ dynamical variables in total). The Hamiltonian system is universally expressible as
\begin{equation}
    \frac{d u_i}{d t} = J_{ij} \frac{\partial H}{\partial u_j},
\end{equation}
where $H$ is the Hamiltonian (a constant of motion), and $J_{ij}$ denotes the elements of the Poisson matrix, whose properties we touch on shortly. This equation can be rewritten in the following fashion:
\begin{equation}\label{P1}
\frac{d u}{d t} = J \frac{\partial H}{\partial u},
\end{equation}
where $u = \left(u_1,u_2,\dots u_N\right)^{T}$ is a vector, and $J$ is the Poisson matrix. To establish that a given dynamical system is Hamiltonian, we must: (1) prove the existence of the Hamiltonian $H$ (which is not necessarily the ``total energy''); and (2) prove that $J$ satisfies the mathematical properties of a noncanonical Poisson bracket, which is defined as
\begin{equation}
    \{f,g\} \equiv \frac{\partial f}{\partial u_i} J_{ij} \frac{\partial g}{\partial u_j}.
\end{equation}
We will now proceed to tackle each of these aspects in turn.

\subsection{Hamiltonian}

In order to determine $H$, we avail ourselves of the fact that $H$ is a constant of motion, implying that we have $dH/dt = 0$. From this fact, we infer that $H$ satisfies the following partial differential equation:
\begin{equation}\label{H}
\frac{d H}{d t}= 0 =\left(\frac{d \alpha_{1}}{d t}\right) \frac{\partial H}{\partial \alpha_{1}}+\left(\frac{d \alpha_{2}}{d t}\right)  \frac{\partial H}{\partial \alpha_{2}}+\left(\frac{d \beta_{1}}{d t}\right)  \frac{\partial H}{\partial \beta_{1}}+\left(\frac{d \beta_{2}}{d t}\right)  \frac{\partial H}{\partial \beta_{2}}+\left(\frac{d b}{d t} \right) \frac{\partial H}{\partial b},
\end{equation}
where $u=(\alpha_{1},\alpha_{2},\beta_{1},\beta_{2}, b)^{T}$ are the dynamical variables of the system. On substituting \eqref{alpha1}-\eqref{b2}, we find that a general solution of (\ref{H}) is 
\begin{equation}\label{H1}
H= -\alpha_{1}\alpha_{2}-d^{2}_{e} \beta_{1}\beta_{2}+\frac{1}{2} d^{2}_{e} b^{2},
\end{equation}
which will serve as our Hamiltonian henceforth. It is observed that the Hamiltonian is a sum of terms of the form $u_i^{n_i}\, u_j^{n_j}$ (recall that the $u$'s are the dynamical variables), where the exponents $n_i$ and $n_j$ are either $0$, $1$, or $2$.

Since we have started from the postulate $dH/dt = 0$ to construct $H$, it is straightforward to verify that (\ref{H1}) does indeed obey the conservation law of $dH/dt = 0$ as originally desired, after using \eqref{alpha1}-\eqref{b2}.

\subsection{Noncanonical Poisson bracket}

The next step in establishing Hamiltonian dynamics is to ascertain the Poisson matrix $J$, and prove that it satisfies the requisite properties of a noncanonical Poisson bracket. 

By combining (\ref{alpha1})-(\ref{b2}) with (\ref{P1}) and the Hamiltonian (\ref{H1}) derived earlier, the candidate Poisson bracket, whose properties need to be evaluated next, is found to have the form:
\begin{equation}\label{Poisson}
J = \begin{pmatrix}
0 & 2(\gamma(t) - d_i b(t)) & 0 & 0 & -2\beta_1(t) \\
2(d_i b(t)-\gamma(t)) & 0 & 0 & 0 & -2\beta_2(t) \\
0 & 0 & 0 & \frac{2}{d_e^2}\gamma(t) & -\frac{2}{d_e^2}\alpha_1(t) \\
0 & 0 & -\frac{2}{d_e^2}\gamma(t) & 0 & -\frac{2}{d_e^2}\alpha_2(t) \\
2\beta_1(t) & 2\beta_2(t) & \frac{2}{d_e^2}\alpha_1(t) & \frac{2}{d_e^2}\alpha_2(t) & 0
\end{pmatrix}
\end{equation}
An immediate point to appreciate is that this Poisson matrix possesses a Lie-Poisson algebraic structure if $\gamma$ is a linear function of the dynamical variables (or is independent of them), which is tantamount to stating that this matrix has an overall \emph{linear} dependence on the dynamical variables \cite{PJM82,PJM98}. This feature is readily verified upon inspection of the above Poisson matrix. However, in this section, we do not explicitly require the generalized Poisson bracket to have this particular structure when analyzing its properties. Recall that the generic version of $\gamma$ was delineated at the end of Section \ref{SecJanda}, whereas we shall tackle a specific ansatz in Section \ref{SSecCasimir}.

Now, we turn our attention to the vital task of demonstrating that (\ref{Poisson}) does possess the requisite features of a noncanonical Poisson bracket. It is well-known that such brackets, also known as generalized Poisson brackets in some sources, must obey the following characteristics (see, e.g., \cite[equation 1.6]{PJM82} and \cite[Chapter 5]{SM16}):

\begin{enumerate}
  \item \textbf{Antisymmetry:}
  \begin{equation}\label{AntiSym}
  J_{ij} = -J_{ji}
  \end{equation}
  In other words, we require the Poisson matrix to be antisymmetric, which is found to be valid upon inspecting (\ref{Poisson}), consequently fulfilling this condition as desired.

  \item \textbf{Jacobi Identity:} \\
  
  As per the Jacobi identity, the following expression $\Sigma_{ijk}$ must sum to zero when computed over all permutations of the indices $i$, $j$, and $k$ \citep[e.g.,][equation 148]{PJM98}.
  \begin{equation}\label{JacobGen}
  \Sigma_{ijk} \equiv \sum_{l=0}^{N-1} J_{li} \frac{\partial J_{jk}}{\partial u_l} + J_{lj} \frac{\partial J_{ki}}{\partial u_l} + J_{lk} \frac{\partial J_{ij}}{\partial u_l}
  \end{equation}
  In other words, the mathematical condition for the Jacobi identity is that
  \begin{equation}\label{Jacobi}
      \sum_{(i,j,k)} \Sigma_{ijk} = 0
  \end{equation}
  The Jacobi identity is a critical condition for ensuring the noncanonical Poisson bracket structure, and thereby bestowing a Hamiltonian structure upon the system.
\end{enumerate}

Since we have already established the antisymmetry of the Poisson bracket, we will tackle the Jacobi identity separately in Section \ref{SSecJacobi}, which is next.

\subsection{Proof of the Jacobi identity}\label{SSecJacobi}

For our system, we recall the notation that was (implicitly) introduced earlier: $u_0 \equiv \alpha_1$, $u_1 \equiv \alpha_2$, $u_2 \equiv \beta_1$, $u_3 \equiv \beta_2$, and $u_4 \equiv b$. To put it differently, the label `$0$' is associated with $\alpha_1$, `$1$' with $\alpha_2$, and so forth.

In this case, we can expand (\ref{JacobGen}) as follows:
\begin{align}\label{SigmaExpfin}
\Sigma_{ijk} &\equiv  \left( J_{0i} \frac{\partial J_{jk}}{\partial \alpha_1} + J_{0j} \frac{\partial J_{ki}}{\partial \alpha_1} + J_{0k} \frac{\partial J_{ij}}{\partial \alpha_1} \right) \\ \nonumber
&+ \left(J_{1i} \frac{\partial J_{jk}}{\partial \alpha_2} + J_{1j} \frac{\partial J_{ki}}{\partial \alpha_2} + J_{1k} \frac{\partial J_{ij}}{\partial \alpha_2} \right) \\ \nonumber
&+  \left( J_{2i} \frac{\partial J_{jk}}{\partial \beta_1} + J_{2j} \frac{\partial J_{ki}}{\partial \beta_1} + J_{2k} \frac{\partial J_{ij}}{\partial \beta_1} \right) \\ \nonumber
&+  \left( J_{3i} \frac{\partial J_{jk}}{\partial \beta_2} + J_{3j} \frac{\partial J_{ki}}{\partial \beta_2} + J_{3k} \frac{\partial J_{ij}}{\partial \beta_2} \right) \\ \nonumber
&+  \left( J_{4i} \frac{\partial J_{jk}}{\partial b} + J_{4j} \frac{\partial J_{ki}}{\partial b} + J_{4k} \frac{\partial J_{ij}}{\partial b} \right)
\end{align}
It is observed that each term in the expression involves the partial derivatives of the elements of \( J \) with respect to the dynamical variables. 

By virtue of the antisymmetry of the Poisson matrix, whenever any two of the indices in $\Sigma_{ijk}$ are equal to each other, we have $\Sigma_{ijk} = 0$. To see why this is the case, let us set $j = k = \zeta$ in (\ref{JacobGen}), which yields
\begin{eqnarray}
    \Sigma_{i \zeta \zeta} &=& \sum_{l=0}^{N-1} J_{li} \frac{\partial J_{\zeta \zeta}}{\partial u_l} + J_{l \zeta} \frac{\partial J_{\zeta i}}{\partial u_l} + J_{l \zeta} \frac{\partial J_{i \zeta}}{\partial u_l} \nonumber \\
    &=& \sum_{l=0}^N J_{l \zeta} \left(\frac{\partial J_{\zeta i}}{\partial u_l} +  \frac{\partial J_{i \zeta}}{\partial u_l} \right) \nonumber \\
    &=& 0,
\end{eqnarray}
where the first term on the RHS of the first line vanished because $J_{\zeta \zeta} = 0$ from (\ref{AntiSym}), and the term inside the parentheses on the second line vanishes after invoking (\ref{AntiSym}). 

Hence, all the remaining potentially nonzero contributions to $\Sigma_{ijk}$ have been explicitly worked out in Appendix \ref{apendixA}. By taking the appropriate sum of these terms in Appendix \ref{apendixA}, it can be verified that (\ref{Jacobi}) is fulfilled, thus completing our proof of the Jacobi identity. In turn, this proof demonstrates that (\ref{Poisson}) constitutes a valid noncanonical Poisson bracket endowed with the necessary mathematical properties.

We have thus established that \eqref{alpha1}-\eqref{b2} is a noncanonical Hamiltonian system, with the Hamiltonian given by (\ref{H1}) and the Poisson matrix by (\ref{Poisson}).

\subsection{Casimir invariants}\label{SSecCasimir}

Noncanonical Poisson brackets are characterized by the existence of a special class of invariants, namely, the Casimir invariants (or Casimirs, for short). The topological significance of the Casimirs is that they serve as invariants of phase space \citep{PJM98,HSS09,YM13,ZY16}. The striking characteristic of Casimirs is that they obey
\begin{equation}
    \{f,C\} = 0 \quad \forall\, f,
\end{equation}
which translates to the condition that $J \cdot \nabla_{u} C = 0$, where \( \nabla_{u} C \) is the gradient of the Casimir invariant, while $J$ is the Poisson matrix given by (\ref{Poisson}).
Therefore, the Casimirs obey
\begin{equation}
    \begin{pmatrix}
0 & 2(\gamma(t) - d_i b(t)) & 0 & 0 & -2\beta_1(t) \\
2(d_i b(t)-\gamma(t)) & 0 & 0 & 0 & -2\beta_2(t) \\
0 & 0 & 0 & \frac{2}{d_e^2}\gamma(t) & -\frac{2}{d_e^2}\alpha_1(t) \\
0 & 0 & -\frac{2}{d_e^2}\gamma(t) & 0 & -\frac{2}{d_e^2}\alpha_2(t) \\
2\beta_1(t) & 2\beta_2(t) & \frac{2}{d_e^2}\alpha_1(t) & \frac{2}{d_e^2}\alpha_2(t) & 0
\end{pmatrix} \begin{pmatrix}
{\partial C}/{\partial \alpha_1} \\
{\partial C}/{\partial \alpha_2} \\
{\partial C}/{\partial \beta_1} \\
{\partial C}/{\partial \beta_2} \\
{\partial C}/{\partial b} \\
\end{pmatrix} = 0,
\end{equation}
which yields the following quintet of partial differential equations:
\begin{align} \label{Set1}
(\gamma(t) - d_i b(t)) \frac{\partial C}{\partial \alpha_2} -  \beta_1(t) \frac{\partial C}{\partial b} &= 0, \\ \label{Set2}
 (d_i b(t) - \gamma(t)) \frac{\partial C}{\partial \alpha_1} -  \beta_2(t) \frac{\partial C}{\partial b} &= 0, \\ \label{Set3}
\gamma(t) \frac{\partial C}{\partial \beta_2} -   \alpha_1(t) \frac{\partial C}{\partial b} &= 0, \\ \label{Set4}
 \gamma(t) \frac{\partial C}{\partial \beta_1} +  \alpha_2(t) \frac{\partial C}{\partial b} &= 0, \\ \label{Set5}
\beta_1(t) \frac{\partial C}{\partial \alpha_1} +  \beta_2(t) \frac{\partial C}{\partial \alpha_2} +   \frac{\alpha_1(t)}{d_e^2} \frac{\partial C}{\partial \beta_1} +  \frac{\alpha_2(t)}{d_e^2} \frac{\partial C}{\partial \beta_2} &= 0.
\end{align}
Although these represent a complicated set of equations, on carrying out the manipulation of (\ref{Set1}) and (\ref{Set2}), we end up with
\begin{equation}\label{IntCond1}
    \frac{\partial C}{\partial \alpha_2} = -\frac{\beta_1}{ \beta_2} \frac{\partial C}{\partial \alpha_1},
\end{equation}
and in a similar fashion, (\ref{Set3}) and (\ref{Set4}) can be combined to obtain
\begin{equation}\label{IntCond2}
    \frac{\partial C}{\partial \beta_2}=-\frac{ \alpha_1}{\alpha_2} \frac{\partial C}{\partial \beta_1}.
\end{equation}
An interesting attribute of (\ref{IntCond1}) and (\ref{IntCond2}) is that they jointly satisfy (\ref{Set5}) automatically. 

At this stage, we remark that solving a system of five nonlinear coupled PDEs to determine the \emph{complete} set of solutions for $C$ is not an easy task, and is consequently left for future work (broadly in the manner of Ref. \cite{TM00}). However, we tackle one particular solution below, with the clear proviso that the condition employed is motivated by mathematical convenience (although it may possess some underlying physical basis, as touched on hereafter). Before doing so, let us recall from the ending of Section \ref{SecJanda} that $\gamma$ seems, in principle, freely specifiable. In a similar vein, we highlight that (\ref{Poisson}) exhibits the requisite mathematical characteristics of a noncanonical Poisson bracket, for arbitrary functional dependence of $\gamma$ on the dynamical variables. Lastly, the Hamiltonian is independent of $\gamma$, as revealed from scrutinizing (\ref{H1}).

On the basis of these useful aspects, we shall adopt one select ansatz for which the ensuing Casimir invariant (\ref{Cas1}) will be shown to exist.
\begin{equation}\label{gammachoice}
    \gamma = \frac{d_i}{2} b,
\end{equation}
motivated by the following considerations:
\begin{itemize}
    \item In \eqref{alpha1} and \eqref{alpha2}, selecting (\ref{gammachoice})  converts the terms inside the parentheses into a more compact form, and also generates a consistent $b$ dependence in \eqref{beta1} and \eqref{beta2}.
    \item The above trend is even more discernible in the Poisson matrix (\ref{Poisson}) if we substitute (\ref{gammachoice}). A crucial point is that (\ref{gammachoice}) ensures the Lie-Poisson structure of (\ref{Poisson}), because all the terms inside the matrix are still linear in the dynamical variables.
    \item Lastly, from a physical standpoint, $b(t)$ is essentially a component of the magnetic field, while $\gamma(t)$ is linearly related to two components of the velocity field (by way of $\phi$). It is well-known that the velocity and magnetic field components of Alfv\'enic waves in incompressible XMHD are linearly proportional to each other, as shown in Refs. \cite{MK05,MM09,AY16,ALM16}. Therefore, by the same token, it is possible that a linear relationship between $\gamma$ and $b$ may be tenable through analogy, along the lines posited in (\ref{gammachoice}).
\end{itemize}

Hence, on proceeding with (\ref{gammachoice}) and \eqref{Set1}-\eqref{Set5}, we find that one of the possible Casimir invariants takes the form:
\begin{equation}\label{Cas1}
    C = \mathcal{C} \left(\beta_2 \alpha_1 - \beta_1 \alpha_2\right) + \mathcal{C} \gamma b,
\end{equation}
where $\mathcal{C}$ represents an arbitrary constant, and $\gamma$ must be understood to satisfy (\ref{gammachoice}). An arbitrary choice of $\gamma$ (which would thus be more general), on the other hand, does not yield this Casimir and solving the full system of nonlinear PDEs for the Casimirs is a challenging task that is well-suited for future research, as outlined previously.

\section{Conclusions}\label{SecConc}

X-point collapse is a widely investigated process in magnetic reconnection. The latter phenomenon is ubiquitous in plasmas, ranging from laboratory and fusion to space and astrophysical plasmas. Magnetic reconnection is often modeled by extended magnetohydrodynamics (XMHD), which has the intrinsic trait of facilitating the breaking of magnetic field lines and permitting fast reconnection. 

In this work, we commence with a Hamiltonian formulation of reduced XMHD -- which is endowed with a suitable Hamiltonian and noncanonical Poisson bracket -- and indicate how it can be reduced to a set of nonlinear coupled ODEs that are presumably capable of describing X-point collapse. This approach entirely parallels the earlier derivation by Ref. \cite{AZJ18}. By starting from a ``parent'' Hamiltonian model, it is expected that its ``offshoot'' would also possess a Hamiltonian formulation.

Motivated by this statement, we seek the noncanonical Hamiltonian structure of the dynamical equations of X-point collapse in the regime of reduced XMHD, which is a problem of direct relevance in plasma physics, as evinced by the cited references in Section \ref{SecIntro}. For this system of five dynamical variables, we demonstrate that a noncanonical Poisson bracket satisfying the necessary conditions does exist, which is given by (\ref{Poisson}). Likewise, we compute a conserved quantity (\ref{H1}), the Hamiltonian of the system. Thus, (\ref{H1}) and (\ref{Poisson}) together comprise the noncanonical Hamiltonian structure of this model. We also compute the governing equations for the Casimir invariants, and thus calculate one specific instance, deferring an exhaustive categorization to future work.

While our analysis can seem formal, we reiterate the connections to plasma physics (viz., the dynamics of X-point collapse) and the benefits of a Hamiltonian formulation sketched at the end of Section \ref{SecIntro}, most notably the fact that devising accurate numerical algorithms is rendered viable, which would be valuable for handling (near-)singularities regularly observed in simulations of magnetic reconnection. It is thus hoped that our work may pave the way for future treatises of this kind.

\begin{appendices}
\section{Explicit proof of the Jacobi identity}\label{apendixA}
\numberwithin{equation}{section}
\setcounter{equation}{0}

Specifically, for all permutations of indices $(i, j, k)$, to demonstrate that the Jacobi identity condition (\ref{Jacobi}) is fulfilled, we must compute the various nonzero terms of $\Sigma_{ijk}$, which are reported here explicitly. We draw on (\ref{SigmaExpfin}) in conjunction with the postulate at the close of Section \ref{SecJanda}, to wit, that $\gamma$ has the general functional form of $\gamma\left(\alpha_1,\alpha_2,\beta_1,\beta_2,b,t\right)$; note that this ansatz also includes the special case where $\gamma$ is independent of the dynamical variables. However, in the bulk of the paper, a particular version of $\gamma$ does not enter any of the equations directly, except for (\ref{gammachoice}) and its following exposition.

\begin{eqnarray}
\Sigma_{012} &=& \bigg( J_{00} \frac{\partial J_{12}}{\partial \alpha_1} + J_{01} \frac{\partial J_{20}}{\partial \alpha_1} + J_{02} \frac{\partial J_{01}}{\partial \alpha_1} + J_{10} \frac{\partial J_{12}}{\partial \alpha_2} + J_{11} \frac{\partial J_{20}}{\partial \alpha_2} + J_{12} \frac{\partial J_{01}}{\partial \alpha_2} + J_{20} \frac{\partial J_{12}}{\partial \beta_1} \\ \nonumber
&+& J_{21} \frac{\partial J_{20}}{\partial \beta_1} + J_{22} \frac{\partial J_{01}}{\partial \beta_1} + J_{30} \frac{\partial J_{12}}{\partial \beta_2} + J_{31} \frac{\partial J_{20}}{\partial \beta_2} + J_{32} \frac{\partial J_{01}}{\partial \beta_2} + J_{40} \frac{\partial J_{12}}{\partial b} + J_{41} \frac{\partial J_{20}}{\partial b} + J_{42} \frac{\partial J_{01}}{\partial b} \bigg)\\ \nonumber
&=&\frac{4\left[-\left(d_i - \frac{\partial \gamma}{\partial b}\right)\alpha_1(t) - \gamma \frac{\partial \gamma}{\partial \beta_2}\right]}{d_e^2}
,
\end{eqnarray}
\begin{eqnarray}
\Sigma_{013} &=& \bigg(J_{00} \frac{\partial J_{13}}{\partial \alpha_1} + J_{01} \frac{\partial J_{30}}{\partial \alpha_1} + J_{03} \frac{\partial J_{01}}{\partial \alpha_1} + J_{10} \frac{\partial J_{13}}{\partial \alpha_2} + J_{11} \frac{\partial J_{30}}{\partial \alpha_2} + J_{13} \frac{\partial J_{01}}{\partial \alpha_2} + J_{20} \frac{\partial J_{13}}{\partial \beta_1} \\ \nonumber
&+& J_{21} \frac{\partial J_{30}}{\partial \beta_1} + J_{23} \frac{\partial J_{01}}{\partial \beta_1} + J_{30} \frac{\partial J_{13}}{\partial \beta_2} + J_{31} \frac{\partial J_{30}}{\partial \beta_2} + J_{33} \frac{\partial J_{01}}{\partial \beta_2} + J_{40} \frac{\partial J_{13}}{\partial b} + J_{41} \frac{\partial J_{30}}{\partial b} + J_{43} \frac{\partial J_{01}}{\partial b} \bigg)\\ \nonumber
&=&\frac{4\left[-\left(d_i - \frac{\partial \gamma}{\partial b}\right)\alpha_2(t) + \gamma \frac{\partial \gamma}{\partial \beta_1}\right]}{d_e^2},
\end{eqnarray}
\begin{eqnarray}
\Sigma_{014} &=& \bigg( J_{00} \frac{\partial J_{14}}{\partial \alpha_1} + J_{01} \frac{\partial J_{40}}{\partial \alpha_1} + J_{04} \frac{\partial J_{01}}{\partial \alpha_1} + J_{10} \frac{\partial J_{14}}{\partial \alpha_2} + J_{11} \frac{\partial J_{40}}{\partial \alpha_2} + J_{14} \frac{\partial J_{01}}{\partial \alpha_2} + J_{20} \frac{\partial J_{14}}{\partial \beta_1}\\ \nonumber
&+& J_{21} \frac{\partial J_{40}}{\partial \beta_1} + J_{24} \frac{\partial J_{01}}{\partial \beta_1} + J_{30} \frac{\partial J_{14}}{\partial \beta_2} + J_{31} \frac{\partial J_{40}}{\partial \beta_2} + J_{34} \frac{\partial J_{01}}{\partial \beta_2} + J_{40} \frac{\partial J_{14}}{\partial b} + J_{41} \frac{\partial J_{40}}{\partial b} + J_{44} \frac{\partial J_{01}}{\partial b} \bigg)\\ \nonumber
&=&\frac{4\left[d_e^2\left(-\beta_1(t)\frac{\partial \gamma}{\partial \alpha_1} - \beta_2(t)\frac{\partial \gamma}{\partial \alpha_2}\right) - \alpha_1(t)\frac{\partial \gamma}{\partial \beta_1} - \alpha_2(t)\frac{\partial \gamma}{\partial \beta_2}\right]}{d_e^2},
\end{eqnarray}
\begin{eqnarray}
\Sigma_{021} &=& \bigg( J_{00} \frac{\partial J_{21}}{\partial \alpha_1} + J_{02} \frac{\partial J_{10}}{\partial \alpha_1} + J_{01} \frac{\partial J_{02}}{\partial \alpha_1} + J_{10} \frac{\partial J_{21}}{\partial \alpha_2} + J_{12} \frac{\partial J_{10}}{\partial \alpha_2} + J_{11} \frac{\partial J_{02}}{\partial \alpha_2} + J_{20} \frac{\partial J_{21}}{\partial \beta_1}\\ \nonumber
& +& J_{22} \frac{\partial J_{10}}{\partial \beta_1} + J_{21} \frac{\partial J_{02}}{\partial \beta_1} + J_{30} \frac{\partial J_{21}}{\partial \beta_2} + J_{32} \frac{\partial J_{10}}{\partial \beta_2} + J_{31} \frac{\partial J_{02}}{\partial \beta_2} + J_{40} \frac{\partial J_{21}}{\partial b} + J_{42} \frac{\partial J_{10}}{\partial b} + J_{41} \frac{\partial J_{02}}{\partial b} \bigg)\\ \nonumber
&=&\frac{4\left[\left(d_i - \frac{\partial \gamma}{\partial b}\right)\alpha_1(t) + \gamma \frac{\partial \gamma}{\partial \beta_2}\right]}{d_e^2},
\end{eqnarray}
\begin{eqnarray}
\Sigma_{023} &=& \bigg(J_{00} \frac{\partial J_{23}}{\partial \alpha_1} + J_{02} \frac{\partial J_{30}}{\partial \alpha_1} + J_{03} \frac{\partial J_{02}}{\partial \alpha_1} + J_{10} \frac{\partial J_{23}}{\partial \alpha_2} + J_{12} \frac{\partial J_{30}}{\partial \alpha_2} + J_{13} \frac{\partial J_{02}}{\partial \alpha_2} + J_{20} \frac{\partial J_{23}}{\partial \beta_1}\\ \nonumber
&+& J_{22} \frac{\partial J_{30}}{\partial \beta_1} + J_{23} \frac{\partial J_{02}}{\partial \beta_1} + J_{30} \frac{\partial J_{23}}{\partial \beta_2} + J_{32} \frac{\partial J_{30}}{\partial \beta_2} + J_{33} \frac{\partial J_{02}}{\partial \beta_2} + J_{40} \frac{\partial J_{23}}{\partial b} + J_{42} \frac{\partial J_{30}}{\partial b} + J_{43} \frac{\partial J_{02}}{\partial b} \bigg)\\ \nonumber
&=&\frac{4\left[\left(d_i b(t) - \gamma\right) \frac{\partial \gamma}{\partial \alpha_2} + \beta_1(t)\frac{\partial \gamma}{\partial b}\right]}{d_e^2},
\end{eqnarray}
\begin{eqnarray}
\Sigma_{024} &=& \bigg( J_{00} \frac{\partial J_{24}}{\partial \alpha_1} + J_{02} \frac{\partial J_{40}}{\partial \alpha_1} + J_{04} \frac{\partial J_{02}}{\partial \alpha_1} + J_{10} \frac{\partial J_{24}}{\partial \alpha_2} + J_{12} \frac{\partial J_{40}}{\partial \alpha_2} + J_{14} \frac{\partial J_{02}}{\partial \alpha_2} + J_{20} \frac{\partial J_{24}}{\partial \beta_1}\\ \nonumber
&+& J_{22} \frac{\partial J_{40}}{\partial \beta_1} + J_{24} \frac{\partial J_{02}}{\partial \beta_1} + J_{30} \frac{\partial J_{24}}{\partial \beta_2} + J_{32} \frac{\partial J_{40}}{\partial \beta_2} + J_{34} \frac{\partial J_{02}}{\partial \beta_2} + J_{40} \frac{\partial J_{24}}{\partial b} + J_{42} \frac{\partial J_{40}}{\partial b} + J_{44} \frac{\partial J_{02}}{\partial b} \bigg)\\ \nonumber
&=&0,
\end{eqnarray}
\begin{eqnarray}
\Sigma_{031} &=& \bigg( J_{00} \frac{\partial J_{31}}{\partial \alpha_1} + J_{03} \frac{\partial J_{10}}{\partial \alpha_1} + J_{01} \frac{\partial J_{03}}{\partial \alpha_1} + J_{10} \frac{\partial J_{31}}{\partial \alpha_2} + J_{13} \frac{\partial J_{10}}{\partial \alpha_2} + J_{11} \frac{\partial J_{03}}{\partial \alpha_2} + J_{20} \frac{\partial J_{31}}{\partial \beta_1}\\ \nonumber
&+& J_{23} \frac{\partial J_{10}}{\partial \beta_1} + J_{21} \frac{\partial J_{03}}{\partial \beta_1} + J_{30} \frac{\partial J_{31}}{\partial \beta_2} + J_{33} \frac{\partial J_{10}}{\partial \beta_2} + J_{31} \frac{\partial J_{03}}{\partial \beta_2} + J_{40} \frac{\partial J_{31}}{\partial b} + J_{43} \frac{\partial J_{10}}{\partial b} + J_{41} \frac{\partial J_{03}}{\partial b} \bigg)\\ \nonumber
&=&\frac{4\left[\left(d_i - \frac{\partial \gamma}{\partial b}\right)\alpha_2(t) - \gamma \frac{\partial \gamma}{\partial \beta_1}\right]}{d_e^2},
\end{eqnarray}
\begin{eqnarray}
\Sigma_{032} &=& \bigg( J_{00} \frac{\partial J_{32}}{\partial \alpha_1} + J_{03} \frac{\partial J_{20}}{\partial \alpha_1} + J_{02} \frac{\partial J_{03}}{\partial \alpha_1} + J_{10} \frac{\partial J_{32}}{\partial \alpha_2} + J_{13} \frac{\partial J_{20}}{\partial \alpha_2} + J_{12} \frac{\partial J_{03}}{\partial \alpha_2} + J_{20} \frac{\partial J_{32}}{\partial \beta_1}\\ \nonumber
&+& J_{23} \frac{\partial J_{20}}{\partial \beta_1} + J_{22} \frac{\partial J_{03}}{\partial \beta_1} + J_{30} \frac{\partial J_{32}}{\partial \beta_2} + J_{33} \frac{\partial J_{20}}{\partial \beta_2} + J_{32} \frac{\partial J_{03}}{\partial \beta_2} + J_{40} \frac{\partial J_{32}}{\partial b} + J_{43} \frac{\partial J_{20}}{\partial b} + J_{42} \frac{\partial J_{03}}{\partial b} \bigg)\\ \nonumber
&=& \frac{4\left[-\left(d_i b(t) - \gamma\right)\frac{\partial \gamma}{\partial \alpha_2} - \beta_1(t)\frac{\partial \gamma}{\partial b}\right]}{d_e^2},
\end{eqnarray}
\begin{eqnarray}
\Sigma_{034} &=& \bigg(J_{00} \frac{\partial J_{34}}{\partial \alpha_1} + J_{03} \frac{\partial J_{40}}{\partial \alpha_1} + J_{04} \frac{\partial J_{03}}{\partial \alpha_1} + J_{10} \frac{\partial J_{34}}{\partial \alpha_2} + J_{13} \frac{\partial J_{40}}{\partial \alpha_2} + J_{14} \frac{\partial J_{03}}{\partial \alpha_2} + J_{20} \frac{\partial J_{34}}{\partial \beta_1}\\ \nonumber
&+& J_{23} \frac{\partial J_{40}}{\partial \beta_1} + J_{24} \frac{\partial J_{03}}{\partial \beta_1} + J_{30} \frac{\partial J_{34}}{\partial \beta_2} + J_{33} \frac{\partial J_{40}}{\partial \beta_2} + J_{34} \frac{\partial J_{03}}{\partial \beta_2} + J_{40} \frac{\partial J_{34}}{\partial b} + J_{43} \frac{\partial J_{40}}{\partial b} + J_{44} \frac{\partial J_{03}}{\partial b} \bigg)\\ \nonumber
&=&\frac{4\left[-d_i b(t) + 2 \gamma\right]}{d_e^2},
\end{eqnarray}
\begin{eqnarray}
\Sigma_{041} &=& \bigg( J_{00} \frac{\partial J_{41}}{\partial \alpha_1} + J_{04} \frac{\partial J_{10}}{\partial \alpha_1} + J_{01} \frac{\partial J_{04}}{\partial \alpha_1} + J_{10} \frac{\partial J_{41}}{\partial \alpha_2} + J_{14} \frac{\partial J_{10}}{\partial \alpha_2} + J_{11} \frac{\partial J_{04}}{\partial \alpha_2} + J_{20} \frac{\partial J_{41}}{\partial \beta_1}\\ \nonumber
&+& J_{24} \frac{\partial J_{10}}{\partial \beta_1} + J_{21} \frac{\partial J_{04}}{\partial \beta_1} + J_{30} \frac{\partial J_{41}}{\partial \beta_2} + J_{34} \frac{\partial J_{10}}{\partial \beta_2} + J_{31} \frac{\partial J_{04}}{\partial \beta_2} + J_{40} \frac{\partial J_{41}}{\partial b} + J_{44} \frac{\partial J_{10}}{\partial b} + J_{41} \frac{\partial J_{04}}{\partial b} \bigg)\\ \nonumber
&=&\frac{4\left[d_e^2\left(\beta_1(t)\frac{\partial \gamma}{\partial \alpha_1} + \beta_2(t)\frac{\partial \gamma}{\partial \alpha_2}\right) + \alpha_1(t)\frac{\partial \gamma}{\partial \beta_1} + \alpha_2(t)\frac{\partial \gamma}{\partial \beta_2}\right]}{d_e^2},
\end{eqnarray}
\begin{eqnarray}
\Sigma_{042} &=& \bigg( J_{00} \frac{\partial J_{42}}{\partial \alpha_1} + J_{04} \frac{\partial J_{20}}{\partial \alpha_1} + J_{02} \frac{\partial J_{04}}{\partial \alpha_1} + J_{10} \frac{\partial J_{42}}{\partial \alpha_2} + J_{14} \frac{\partial J_{20}}{\partial \alpha_2} + J_{12} \frac{\partial J_{04}}{\partial \alpha_2} + J_{20} \frac{\partial J_{42}}{\partial \beta_1}\\ \nonumber
&& J_{24} \frac{\partial J_{20}}{\partial \beta_1} + J_{22} \frac{\partial J_{04}}{\partial \beta_1} + J_{30} \frac{\partial J_{42}}{\partial \beta_2} + J_{34} \frac{\partial J_{20}}{\partial \beta_2} + J_{32} \frac{\partial J_{04}}{\partial \beta_2} + J_{40} \frac{\partial J_{42}}{\partial b} + J_{44} \frac{\partial J_{20}}{\partial b} + J_{42} \frac{\partial J_{04}}{\partial b} \bigg)\\ \nonumber
&=&0,
\end{eqnarray}
\begin{eqnarray}
\Sigma_{043} &=& \bigg( J_{00} \frac{\partial J_{43}}{\partial \alpha_1} + J_{04} \frac{\partial J_{30}}{\partial \alpha_1} + J_{03} \frac{\partial J_{04}}{\partial \alpha_1} + J_{10} \frac{\partial J_{43}}{\partial \alpha_2} + J_{14} \frac{\partial J_{30}}{\partial \alpha_2} + J_{13} \frac{\partial J_{04}}{\partial \alpha_2} + J_{20} \frac{\partial J_{43}}{\partial \beta_1}\\ \nonumber
&+& J_{24} \frac{\partial J_{30}}{\partial \beta_1} + J_{23} \frac{\partial J_{04}}{\partial \beta_1} + J_{30} \frac{\partial J_{43}}{\partial \beta_2} + J_{34} \frac{\partial J_{30}}{\partial \beta_2} + J_{33} \frac{\partial J_{04}}{\partial \beta_2} + J_{40} \frac{\partial J_{43}}{\partial b} + J_{44} \frac{\partial J_{30}}{\partial b} + J_{43} \frac{\partial J_{04}}{\partial b} \bigg)\\ \nonumber
&=&\frac{4\left[d_i b(t) - 2 \gamma\right]}{d_e^2},
\end{eqnarray}
\begin{eqnarray}
\Sigma_{102} &=& \bigg( J_{01} \frac{\partial J_{02}}{\partial \alpha_1} + J_{00} \frac{\partial J_{21}}{\partial \alpha_1} + J_{02} \frac{\partial J_{10}}{\partial \alpha_1} + J_{11} \frac{\partial J_{02}}{\partial \alpha_2} + J_{10} \frac{\partial J_{21}}{\partial \alpha_2} + J_{12} \frac{\partial J_{10}}{\partial \alpha_2} + J_{21} \frac{\partial J_{02}}{\partial \beta_1}\\ \nonumber
&+& J_{20} \frac{\partial J_{21}}{\partial \beta_1} + J_{22} \frac{\partial J_{10}}{\partial \beta_1} + J_{31} \frac{\partial J_{02}}{\partial \beta_2} + J_{30} \frac{\partial J_{21}}{\partial \beta_2} + J_{32} \frac{\partial J_{10}}{\partial \beta_2} + J_{41} \frac{\partial J_{02}}{\partial b} + J_{40} \frac{\partial J_{21}}{\partial b} + J_{42} \frac{\partial J_{10}}{\partial b} \bigg)\\ \nonumber
&=&\frac{4\left[\left(d_i - \frac{\partial \gamma}{\partial b}\right)\alpha_1(t) + \gamma \frac{\partial \gamma}{\partial \beta_2}\right]}{d_e^2},
\end{eqnarray}
\begin{eqnarray}
\Sigma_{103} &=& \bigg( J_{01} \frac{\partial J_{03}}{\partial \alpha_1} + J_{00} \frac{\partial J_{31}}{\partial \alpha_1} + J_{03} \frac{\partial J_{10}}{\partial \alpha_1} + J_{11} \frac{\partial J_{03}}{\partial \alpha_2} + J_{10} \frac{\partial J_{31}}{\partial \alpha_2} + J_{13} \frac{\partial J_{10}}{\partial \alpha_2} + J_{21} \frac{\partial J_{03}}{\partial \beta_1} \\ \nonumber 
&+& J_{20} \frac{\partial J_{31}}{\partial \beta_1} + J_{23} \frac{\partial J_{10}}{\partial \beta_1} + J_{31} \frac{\partial J_{03}}{\partial \beta_2} + J_{30} \frac{\partial J_{31}}{\partial \beta_2} + J_{33} \frac{\partial J_{10}}{\partial \beta_2} + J_{41} \frac{\partial J_{03}}{\partial b} + J_{40} \frac{\partial J_{31}}{\partial b} + J_{43} \frac{\partial J_{10}}{\partial b} \bigg)\\ \nonumber
&=&\frac{4\left[\left(d_i - \frac{\partial \gamma}{\partial b}\right)\alpha_2(t) - \gamma \frac{\partial \gamma}{\partial \beta_1}\right]}{d_e^2},
\end{eqnarray}
\begin{eqnarray}
\Sigma_{104} &=& \bigg( J_{01} \frac{\partial J_{04}}{\partial \alpha_1} + J_{00} \frac{\partial J_{41}}{\partial \alpha_1} + J_{04} \frac{\partial J_{10}}{\partial \alpha_1} + J_{11} \frac{\partial J_{04}}{\partial \alpha_2} + J_{10} \frac{\partial J_{41}}{\partial \alpha_2} + J_{14} \frac{\partial J_{10}}{\partial \alpha_2} + J_{21} \frac{\partial J_{04}}{\partial \beta_1}\\ \nonumber
&+& J_{20} \frac{\partial J_{41}}{\partial \beta_1} + J_{24} \frac{\partial J_{10}}{\partial \beta_1} + J_{31} \frac{\partial J_{04}}{\partial \beta_2} + J_{30} \frac{\partial J_{41}}{\partial \beta_2} + J_{34} \frac{\partial J_{10}}{\partial \beta_2} + J_{41} \frac{\partial J_{04}}{\partial b} + J_{40} \frac{\partial J_{41}}{\partial b} + J_{44} \frac{\partial J_{10}}{\partial b} \bigg)\\ \nonumber
&=&\frac{4\left[d_e^2\left(\beta_1(t)\frac{\partial \gamma}{\partial \alpha_1} + \beta_2(t)\frac{\partial \gamma}{\partial \alpha_2}\right) + \alpha_1(t)\frac{\partial \gamma}{\partial \beta_1} + \alpha_2(t)\frac{\partial \gamma}{\partial \beta_2}\right]}{d_e^2},
\end{eqnarray}
\begin{eqnarray}
\Sigma_{120} &=& \bigg( J_{01} \frac{\partial J_{20}}{\partial \alpha_1} + J_{02} \frac{\partial J_{01}}{\partial \alpha_1} + J_{00} \frac{\partial J_{12}}{\partial \alpha_1} + J_{11} \frac{\partial J_{20}}{\partial \alpha_2} + J_{12} \frac{\partial J_{01}}{\partial \alpha_2} + J_{10} \frac{\partial J_{12}}{\partial \alpha_2} + J_{21} \frac{\partial J_{20}}{\partial \beta_1}\\ \nonumber  
&+& J_{22} \frac{\partial J_{01}}{\partial \beta_1} + J_{20} \frac{\partial J_{12}}{\partial \beta_1} + J_{31} \frac{\partial J_{20}}{\partial \beta_2} + J_{32} \frac{\partial J_{01}}{\partial \beta_2} + J_{30} \frac{\partial J_{12}}{\partial \beta_2} + J_{41} \frac{\partial J_{20}}{\partial b} + J_{42} \frac{\partial J_{01}}{\partial b} + J_{40} \frac{\partial J_{12}}{\partial b} \bigg)\\ \nonumber
&=& \frac{4\left[-\left(d_i - \frac{\partial \gamma}{\partial b}\right)\alpha_1(t) - \gamma \frac{\partial \gamma}{\partial \beta_2}\right]}{d_e^2},
\end{eqnarray}
\begin{eqnarray}
\Sigma_{123} &=& \bigg( J_{01} \frac{\partial J_{23}}{\partial \alpha_1} + J_{02} \frac{\partial J_{31}}{\partial \alpha_1} + J_{03} \frac{\partial J_{12}}{\partial \alpha_1} + J_{11} \frac{\partial J_{23}}{\partial \alpha_2} + J_{12} \frac{\partial J_{31}}{\partial \alpha_2} + J_{13} \frac{\partial J_{12}}{\partial \alpha_2} + J_{21} \frac{\partial J_{23}}{\partial \beta_1}\\ \nonumber
&+& J_{22} \frac{\partial J_{31}}{\partial \beta_1} + J_{23} \frac{\partial J_{12}}{\partial \beta_1} + J_{31} \frac{\partial J_{23}}{\partial \beta_2} + J_{32} \frac{\partial J_{31}}{\partial \beta_2} + J_{33} \frac{\partial J_{12}}{\partial \beta_2} + J_{41} \frac{\partial J_{23}}{\partial b} + J_{42} \frac{\partial J_{31}}{\partial b} + J_{43} \frac{\partial J_{12}}{\partial b} \bigg)\\ \nonumber
&=&\frac{4\left[-\left(d_i b(t) - \gamma\right)\frac{\partial \gamma}{\partial \alpha_1} + \beta_2(t)\frac{\partial \gamma}{\partial b}\right]}{d_e^2},
\end{eqnarray}
\begin{eqnarray}
\Sigma_{124} &=& \bigg( J_{01} \frac{\partial J_{24}}{\partial \alpha_1} + J_{02} \frac{\partial J_{41}}{\partial \alpha_1} + J_{04} \frac{\partial J_{12}}{\partial \alpha_1} + J_{11} \frac{\partial J_{24}}{\partial \alpha_2} + J_{12} \frac{\partial J_{41}}{\partial \alpha_2} + J_{14} \frac{\partial J_{12}}{\partial \alpha_2} + J_{21} \frac{\partial J_{24}}{\partial \beta_1}\\ \nonumber
&+& J_{22} \frac{\partial J_{41}}{\partial \beta_1} + J_{24} \frac{\partial J_{12}}{\partial \beta_1} + J_{31} \frac{\partial J_{24}}{\partial \beta_2} + J_{32} \frac{\partial J_{41}}{\partial \beta_2} + J_{34} \frac{\partial J_{12}}{\partial \beta_2} + J_{41} \frac{\partial J_{24}}{\partial b} + J_{42} \frac{\partial J_{41}}{\partial b} + J_{44} \frac{\partial J_{12}}{\partial b} \bigg)\\ \nonumber
&=&\frac{4\left[d_i b(t) - 2 \gamma\right]}{d_e^2},
\end{eqnarray}
\begin{eqnarray}
\Sigma_{130} &=& \bigg( J_{01} \frac{\partial J_{30}}{\partial \alpha_1} + J_{03} \frac{\partial J_{01}}{\partial \alpha_1} + J_{00} \frac{\partial J_{13}}{\partial \alpha_1} + J_{11} \frac{\partial J_{30}}{\partial \alpha_2} + J_{13} \frac{\partial J_{01}}{\partial \alpha_2} + J_{10} \frac{\partial J_{13}}{\partial \alpha_2} + J_{21} \frac{\partial J_{30}}{\partial \beta_1}\\ \nonumber 
&+& J_{23} \frac{\partial J_{01}}{\partial \beta_1} + J_{20} \frac{\partial J_{13}}{\partial \beta_1} + J_{31} \frac{\partial J_{30}}{\partial \beta_2} + J_{33} \frac{\partial J_{01}}{\partial \beta_2} + J_{30} \frac{\partial J_{13}}{\partial \beta_2} + J_{41} \frac{\partial J_{30}}{\partial b} + J_{43} \frac{\partial J_{01}}{\partial b} + J_{40} \frac{\partial J_{13}}{\partial b} \bigg)\\ \nonumber
&=&\frac{4\left[-\left(d_i - \frac{\partial \gamma}{\partial b}\right)\alpha_2(t) + \gamma \frac{\partial \gamma}{\partial \beta_1}\right]}{d_e^2},
\end{eqnarray}
\begin{eqnarray}
\Sigma_{132} &=& \bigg( J_{01} \frac{\partial J_{32}}{\partial \alpha_1} + J_{03} \frac{\partial J_{21}}{\partial \alpha_1} + J_{02} \frac{\partial J_{13}}{\partial \alpha_1} + J_{11} \frac{\partial J_{32}}{\partial \alpha_2} + J_{13} \frac{\partial J_{21}}{\partial \alpha_2} + J_{12} \frac{\partial J_{13}}{\partial \alpha_2} + J_{21} \frac{\partial J_{32}}{\partial \beta_1}\\ \nonumber
&+& J_{23} \frac{\partial J_{21}}{\partial \beta_1} + J_{22} \frac{\partial J_{13}}{\partial \beta_1} + J_{31} \frac{\partial J_{32}}{\partial \beta_2} + J_{33} \frac{\partial J_{21}}{\partial \beta_2} + J_{32} \frac{\partial J_{13}}{\partial \beta_2} + J_{41} \frac{\partial J_{32}}{\partial b} + J_{43} \frac{\partial J_{21}}{\partial b} + J_{42} \frac{\partial J_{13}}{\partial b} \bigg)\\ \nonumber
&=&\frac{4\left[\left(d_i b(t) - \gamma\right)\frac{\partial \gamma}{\partial \alpha_1} - \beta_2(t)\frac{\partial \gamma}{\partial b}\right]}{d_e^2},
\end{eqnarray}
\begin{eqnarray}
\Sigma_{134} &=& \bigg( J_{01} \frac{\partial J_{34}}{\partial \alpha_1} + J_{03} \frac{\partial J_{41}}{\partial \alpha_1} + J_{04} \frac{\partial J_{13}}{\partial \alpha_1} + J_{11} \frac{\partial J_{34}}{\partial \alpha_2} + J_{13} \frac{\partial J_{41}}{\partial \alpha_2} + J_{14} \frac{\partial J_{13}}{\partial \alpha_2} + J_{21} \frac{\partial J_{34}}{\partial \beta_1}\\ \nonumber
&+& J_{23} \frac{\partial J_{41}}{\partial \beta_1} + J_{24} \frac{\partial J_{13}}{\partial \beta_1} + J_{31} \frac{\partial J_{34}}{\partial \beta_2} + J_{33} \frac{\partial J_{41}}{\partial \beta_2} + J_{34} \frac{\partial J_{13}}{\partial \beta_2} + J_{41} \frac{\partial J_{34}}{\partial b} + J_{43} \frac{\partial J_{41}}{\partial b} + J_{44} \frac{\partial J_{13}}{\partial b} \bigg)\\ \nonumber
&=&0,
\end{eqnarray}
\begin{eqnarray}
\Sigma_{140} &=& \bigg( J_{01} \frac{\partial J_{40}}{\partial \alpha_1} + J_{04} \frac{\partial J_{01}}{\partial \alpha_1} + J_{00} \frac{\partial J_{14}}{\partial \alpha_1} + J_{11} \frac{\partial J_{40}}{\partial \alpha_2} + J_{14} \frac{\partial J_{01}}{\partial \alpha_2} + J_{10} \frac{\partial J_{14}}{\partial \alpha_2} + J_{21} \frac{\partial J_{40}}{\partial \beta_1}\\ \nonumber
&+& J_{24} \frac{\partial J_{01}}{\partial \beta_1} + J_{20} \frac{\partial J_{14}}{\partial \beta_1} + J_{31} \frac{\partial J_{40}}{\partial \beta_2} + J_{34} \frac{\partial J_{01}}{\partial \beta_2} + J_{30} \frac{\partial J_{14}}{\partial \beta_2} + J_{41} \frac{\partial J_{40}}{\partial b} + J_{44} \frac{\partial J_{01}}{\partial b} + J_{40} \frac{\partial J_{14}}{\partial b} \bigg)\\ \nonumber
&=& \frac{4\left[d_e^2\left(-\beta_1(t)\frac{\partial \gamma}{\partial \alpha_1} - \beta_2(t)\frac{\partial \gamma}{\partial \alpha_2}\right) - \alpha_1(t)\frac{\partial \gamma}{\partial \beta_1} - \alpha_2(t)\frac{\partial \gamma}{\partial \beta_2}\right]}{d_e^2},
\end{eqnarray}
\begin{eqnarray}
\Sigma_{142} &=& \bigg( J_{01} \frac{\partial J_{42}}{\partial \alpha_1} + J_{04} \frac{\partial J_{21}}{\partial \alpha_1} + J_{02} \frac{\partial J_{14}}{\partial \alpha_1} + J_{11} \frac{\partial J_{42}}{\partial \alpha_2} + J_{14} \frac{\partial J_{21}}{\partial \alpha_2} + J_{12} \frac{\partial J_{14}}{\partial \alpha_2} + J_{21} \frac{\partial J_{42}}{\partial \beta_1}\\ \nonumber 
&+& J_{24} \frac{\partial J_{21}}{\partial \beta_1} + J_{22} \frac{\partial J_{14}}{\partial \beta_1} + J_{31} \frac{\partial J_{42}}{\partial \beta_2} + J_{34} \frac{\partial J_{21}}{\partial \beta_2} + J_{32} \frac{\partial J_{14}}{\partial \beta_2} + J_{41} \frac{\partial J_{42}}{\partial b} + J_{44} \frac{\partial J_{21}}{\partial b} + J_{42} \frac{\partial J_{14}}{\partial b} \bigg)\\ \nonumber
&=& \frac{4\left[-d_i b(t) + 2 \gamma\right]}{d_e^2},
\end{eqnarray}
\begin{eqnarray}
\Sigma_{143} &=& \bigg( J_{01} \frac{\partial J_{43}}{\partial \alpha_1} + J_{04} \frac{\partial J_{31}}{\partial \alpha_1} + J_{03} \frac{\partial J_{14}}{\partial \alpha_1} + J_{11} \frac{\partial J_{43}}{\partial \alpha_2} + J_{14} \frac{\partial J_{31}}{\partial \alpha_2} + J_{13} \frac{\partial J_{14}}{\partial \alpha_2} + J_{21} \frac{\partial J_{43}}{\partial \beta_1}\\ \nonumber 
&+& J_{24} \frac{\partial J_{31}}{\partial \beta_1} + J_{23} \frac{\partial J_{14}}{\partial \beta_1} + J_{31} \frac{\partial J_{43}}{\partial \beta_2} + J_{34} \frac{\partial J_{31}}{\partial \beta_2} + J_{33} \frac{\partial J_{14}}{\partial \beta_2} + J_{41} \frac{\partial J_{43}}{\partial b} + J_{44} \frac{\partial J_{31}}{\partial b} + J_{43} \frac{\partial J_{14}}{\partial b} \bigg)\\ \nonumber
&=&0,
\end{eqnarray}
\begin{eqnarray}
\Sigma_{201} &=& \bigg( J_{02} \frac{\partial J_{01}}{\partial \alpha_1} + J_{00} \frac{\partial J_{12}}{\partial \alpha_1} + J_{01} \frac{\partial J_{20}}{\partial \alpha_1} + J_{12} \frac{\partial J_{01}}{\partial \alpha_2} + J_{10} \frac{\partial J_{12}}{\partial \alpha_2} + J_{11} \frac{\partial J_{20}}{\partial \alpha_2} + J_{22} \frac{\partial J_{01}}{\partial \beta_1}\\ \nonumber
&+& J_{20} \frac{\partial J_{12}}{\partial \beta_1} + J_{21} \frac{\partial J_{20}}{\partial \beta_1} + J_{32} \frac{\partial J_{01}}{\partial \beta_2} + J_{30} \frac{\partial J_{12}}{\partial \beta_2} + J_{31} \frac{\partial J_{20}}{\partial \beta_2} + J_{42} \frac{\partial J_{01}}{\partial b} + J_{40} \frac{\partial J_{12}}{\partial b} + J_{41} \frac{\partial J_{20}}{\partial b} \bigg)\\ \nonumber
&=& \frac{4\left[-\left(d_i - \frac{\partial \gamma}{\partial b}\right)\alpha_1(t) - \gamma \frac{\partial \gamma}{\partial \beta_2}\right]}{d_e^2},
\end{eqnarray}
\begin{eqnarray}
\Sigma_{203} &=& \bigg( J_{02} \frac{\partial J_{03}}{\partial \alpha_1} + J_{00} \frac{\partial J_{32}}{\partial \alpha_1} + J_{03} \frac{\partial J_{20}}{\partial \alpha_1} + J_{12} \frac{\partial J_{03}}{\partial \alpha_2} + J_{10} \frac{\partial J_{32}}{\partial \alpha_2} + J_{13} \frac{\partial J_{20}}{\partial \alpha_2} + J_{22} \frac{\partial J_{03}}{\partial \beta_1}\\ \nonumber
&+& J_{20} \frac{\partial J_{32}}{\partial \beta_1} + J_{23} \frac{\partial J_{20}}{\partial \beta_1} + J_{32} \frac{\partial J_{03}}{\partial \beta_2} + J_{30} \frac{\partial J_{32}}{\partial \beta_2} + J_{33} \frac{\partial J_{20}}{\partial \beta_2} + J_{42} \frac{\partial J_{03}}{\partial b} + J_{40} \frac{\partial J_{32}}{\partial b} + J_{43} \frac{\partial J_{20}}{\partial b} \bigg)\\ \nonumber
&=&\frac{4\left[-\left(d_i b(t) - \gamma\right)\frac{\partial \gamma}{\partial \alpha_2} - \beta_1(t)\frac{\partial \gamma}{\partial b}\right]}{d_e^2},
\end{eqnarray}
\begin{eqnarray}
\Sigma_{204} &=& \bigg( J_{02} \frac{\partial J_{04}}{\partial \alpha_1} + J_{00} \frac{\partial J_{42}}{\partial \alpha_1} + J_{04} \frac{\partial J_{20}}{\partial \alpha_1} + J_{12} \frac{\partial J_{04}}{\partial \alpha_2} + J_{10} \frac{\partial J_{42}}{\partial \alpha_2} + J_{14} \frac{\partial J_{20}}{\partial \alpha_2} + J_{22} \frac{\partial J_{04}}{\partial \beta_1}\\ \nonumber
&+& J_{20} \frac{\partial J_{42}}{\partial \beta_1} + J_{24} \frac{\partial J_{20}}{\partial \beta_1} + J_{32} \frac{\partial J_{04}}{\partial \beta_2} + J_{30} \frac{\partial J_{42}}{\partial \beta_2} + J_{34} \frac{\partial J_{20}}{\partial \beta_2} + J_{42} \frac{\partial J_{04}}{\partial b} + J_{40} \frac{\partial J_{42}}{\partial b} + J_{44} \frac{\partial J_{20}}{\partial b} \bigg)\\ \nonumber
&=&0,
\end{eqnarray}
\begin{eqnarray}
\Sigma_{210} &=& \bigg( J_{02} \frac{\partial J_{10}}{\partial \alpha_1} + J_{01} \frac{\partial J_{02}}{\partial \alpha_1} + J_{00} \frac{\partial J_{21}}{\partial \alpha_1} + J_{12} \frac{\partial J_{10}}{\partial \alpha_2} + J_{11} \frac{\partial J_{02}}{\partial \alpha_2} + J_{10} \frac{\partial J_{21}}{\partial \alpha_2} + J_{22} \frac{\partial J_{10}}{\partial \beta_1}\\ \nonumber
&+& J_{21} \frac{\partial J_{02}}{\partial \beta_1} + J_{20} \frac{\partial J_{21}}{\partial \beta_1} + J_{32} \frac{\partial J_{10}}{\partial \beta_2} + J_{31} \frac{\partial J_{02}}{\partial \beta_2} + J_{30} \frac{\partial J_{21}}{\partial \beta_2} + J_{42} \frac{\partial J_{10}}{\partial b} + J_{41} \frac{\partial J_{02}}{\partial b} + J_{40} \frac{\partial J_{21}}{\partial b} \bigg)\\ \nonumber
&=&\frac{4\left[\left(d_i - \frac{\partial \gamma}{\partial b}\right)\alpha_1(t) + \gamma \frac{\partial \gamma}{\partial \beta_2}\right]}{d_e^2},
\end{eqnarray}
\begin{eqnarray}
\Sigma_{213} &=& \bigg( J_{02} \frac{\partial J_{13}}{\partial \alpha_1} + J_{01} \frac{\partial J_{32}}{\partial \alpha_1} + J_{03} \frac{\partial J_{21}}{\partial \alpha_1} + J_{12} \frac{\partial J_{13}}{\partial \alpha_2} + J_{11} \frac{\partial J_{32}}{\partial \alpha_2} + J_{13} \frac{\partial J_{21}}{\partial \alpha_2} + J_{22} \frac{\partial J_{13}}{\partial \beta_1}\\ \nonumber
&+& J_{21} \frac{\partial J_{32}}{\partial \beta_1} + J_{23} \frac{\partial J_{21}}{\partial \beta_1} + J_{32} \frac{\partial J_{13}}{\partial \beta_2} + J_{31} \frac{\partial J_{32}}{\partial \beta_2} + J_{33} \frac{\partial J_{21}}{\partial \beta_2} + J_{42} \frac{\partial J_{13}}{\partial b} + J_{41} \frac{\partial J_{32}}{\partial b} + J_{43} \frac{\partial J_{21}}{\partial b} \bigg)\\ \nonumber
&=&\frac{4\left[\left(d_i b(t) - \gamma\right)\frac{\partial \gamma}{\partial \alpha_1} - \beta_2(t)\frac{\partial \gamma}{\partial b}\right]}{d_e^2},
\end{eqnarray}
\begin{eqnarray}
\Sigma_{214} &=& \bigg( J_{02} \frac{\partial J_{14}}{\partial \alpha_1} + J_{01} \frac{\partial J_{42}}{\partial \alpha_1} + J_{04} \frac{\partial J_{21}}{\partial \alpha_1} + J_{12} \frac{\partial J_{14}}{\partial \alpha_2} + J_{11} \frac{\partial J_{42}}{\partial \alpha_2} + J_{14} \frac{\partial J_{21}}{\partial \alpha_2} + J_{22} \frac{\partial J_{14}}{\partial \beta_1}\\ \nonumber
&+& J_{21} \frac{\partial J_{42}}{\partial \beta_1} + J_{24} \frac{\partial J_{21}}{\partial \beta_1} + J_{32} \frac{\partial J_{14}}{\partial \beta_2} + J_{31} \frac{\partial J_{42}}{\partial \beta_2} + J_{34} \frac{\partial J_{21}}{\partial \beta_2} + J_{42} \frac{\partial J_{14}}{\partial b} + J_{41} \frac{\partial J_{42}}{\partial b} + J_{44} \frac{\partial J_{21}}{\partial b} \bigg)\\ \nonumber
&=&\frac{4\left[-d_i b(t)+ 2 \gamma\right]}{d_e^2},
\end{eqnarray}
\begin{eqnarray}
\Sigma_{230} &=& \bigg( J_{02} \frac{\partial J_{30}}{\partial \alpha_1} + J_{03} \frac{\partial J_{02}}{\partial \alpha_1} + J_{00} \frac{\partial J_{23}}{\partial \alpha_1} + J_{12} \frac{\partial J_{30}}{\partial \alpha_2} + J_{13} \frac{\partial J_{02}}{\partial \alpha_2} + J_{10} \frac{\partial J_{23}}{\partial \alpha_2} + J_{22} \frac{\partial J_{30}}{\partial \beta_1}\\ \nonumber
&+& J_{23} \frac{\partial J_{02}}{\partial \beta_1} + J_{20} \frac{\partial J_{23}}{\partial \beta_1} + J_{32} \frac{\partial J_{30}}{\partial \beta_2} + J_{33} \frac{\partial J_{02}}{\partial \beta_2} + J_{30} \frac{\partial J_{23}}{\partial \beta_2} + J_{42} \frac{\partial J_{30}}{\partial b} + J_{43} \frac{\partial J_{02}}{\partial b} + J_{40} \frac{\partial J_{23}}{\partial b} \bigg)\\ \nonumber
&=&\frac{4\left[\left(d_i b(t) - \gamma\right)\frac{\partial \gamma}{\partial \alpha_2} + \beta_1(t)\frac{\partial \gamma}{\partial b}\right]}{d_e^2},
\end{eqnarray}
\begin{eqnarray}
\Sigma_{231} &=& \bigg( J_{02} \frac{\partial J_{31}}{\partial \alpha_1} + J_{03} \frac{\partial J_{12}}{\partial \alpha_1} + J_{01} \frac{\partial J_{23}}{\partial \alpha_1} + J_{12} \frac{\partial J_{31}}{\partial \alpha_2} + J_{13} \frac{\partial J_{12}}{\partial \alpha_2} + J_{11} \frac{\partial J_{23}}{\partial \alpha_2} + J_{22} \frac{\partial J_{31}}{\partial \beta_1}\\ \nonumber
&+& J_{23} \frac{\partial J_{12}}{\partial \beta_1} + J_{21} \frac{\partial J_{23}}{\partial \beta_1} + J_{32} \frac{\partial J_{31}}{\partial \beta_2} + J_{33} \frac{\partial J_{12}}{\partial \beta_2} + J_{31} \frac{\partial J_{23}}{\partial \beta_2} + J_{42} \frac{\partial J_{31}}{\partial b} + J_{43} \frac{\partial J_{12}}{\partial b} + J_{41} \frac{\partial J_{23}}{\partial b} \bigg)\\ \nonumber
&=& \frac{4\left[-\left(d_i b(t) - \gamma\right)\frac{\partial \gamma}{\partial \alpha_1} + \beta_2(t)\frac{\partial \gamma}{\partial b}\right]}{d_e^2},
\end{eqnarray}
\begin{eqnarray}
\Sigma_{234} &=& \bigg( J_{02} \frac{\partial J_{34}}{\partial \alpha_1} + J_{03} \frac{\partial J_{42}}{\partial \alpha_1} + J_{04} \frac{\partial J_{23}}{\partial \alpha_1} + J_{12} \frac{\partial J_{34}}{\partial \alpha_2} + J_{13} \frac{\partial J_{42}}{\partial \alpha_2} + J_{14} \frac{\partial J_{23}}{\partial \alpha_2} + J_{22} \frac{\partial J_{34}}{\partial \beta_1}\\ \nonumber
&+& J_{23} \frac{\partial J_{42}}{\partial \beta_1} + J_{24} \frac{\partial J_{23}}{\partial \beta_1} + J_{32} \frac{\partial J_{34}}{\partial \beta_2} + J_{33} \frac{\partial J_{42}}{\partial \beta_2} + J_{34} \frac{\partial J_{23}}{\partial \beta_2} + J_{42} \frac{\partial J_{34}}{\partial b} + J_{43} \frac{\partial J_{42}}{\partial b} + J_{44} \frac{\partial J_{23}}{\partial b} \bigg)\\ \nonumber
&=&\frac{4\left[d_e^2\left(-\beta_1(t)\frac{\partial \gamma}{\partial \alpha_1} - \beta_2(t)\frac{\partial \gamma}{\partial \alpha_2}\right) - \alpha_1(t)\frac{\partial \gamma}{\partial \beta_1} - \alpha_2(t)\frac{\partial \gamma}{\partial \beta_2}\right]}{d_e^4},
\end{eqnarray}
\begin{eqnarray}
\Sigma_{240} &=& \bigg( J_{02} \frac{\partial J_{40}}{\partial \alpha_1} + J_{04} \frac{\partial J_{02}}{\partial \alpha_1} + J_{00} \frac{\partial J_{24}}{\partial \alpha_1} + J_{12} \frac{\partial J_{40}}{\partial \alpha_2} + J_{14} \frac{\partial J_{02}}{\partial \alpha_2} + J_{10} \frac{\partial J_{24}}{\partial \alpha_2} + J_{22} \frac{\partial J_{40}}{\partial \beta_1}\\ \nonumber
&+& J_{24} \frac{\partial J_{02}}{\partial \beta_1} + J_{20} \frac{\partial J_{24}}{\partial \beta_1} + J_{32} \frac{\partial J_{40}}{\partial \beta_2} + J_{34} \frac{\partial J_{02}}{\partial \beta_2} + J_{30} \frac{\partial J_{24}}{\partial \beta_2} + J_{42} \frac{\partial J_{40}}{\partial b} + J_{44} \frac{\partial J_{02}}{\partial b} + J_{40} \frac{\partial J_{24}}{\partial b} \bigg)\\ \nonumber
&=&0,
\end{eqnarray}
\begin{eqnarray}
\Sigma_{241} &=& \bigg( J_{02} \frac{\partial J_{41}}{\partial \alpha_1} + J_{04} \frac{\partial J_{12}}{\partial \alpha_1} + J_{01} \frac{\partial J_{24}}{\partial \alpha_1} + J_{12} \frac{\partial J_{41}}{\partial \alpha_2} + J_{14} \frac{\partial J_{12}}{\partial \alpha_2} + J_{11} \frac{\partial J_{24}}{\partial \alpha_2} + J_{22} \frac{\partial J_{41}}{\partial \beta_1}\\ \nonumber
&+& J_{24} \frac{\partial J_{12}}{\partial \beta_1} + J_{21} \frac{\partial J_{24}}{\partial \beta_1} + J_{32} \frac{\partial J_{41}}{\partial \beta_2} + J_{34} \frac{\partial J_{12}}{\partial \beta_2} + J_{31} \frac{\partial J_{24}}{\partial \beta_2} + J_{42} \frac{\partial J_{41}}{\partial b} + J_{44} \frac{\partial J_{12}}{\partial b} + J_{41} \frac{\partial J_{24}}{\partial b} \bigg)\\ \nonumber
&=&\frac{4\left[d_i b(t) - 2 \gamma\right]}{d_e^2},
\end{eqnarray}
\begin{eqnarray}
\Sigma_{243} &=& \bigg( J_{02} \frac{\partial J_{43}}{\partial \alpha_1} + J_{04} \frac{\partial J_{32}}{\partial \alpha_1} + J_{03} \frac{\partial J_{24}}{\partial \alpha_1} + J_{12} \frac{\partial J_{43}}{\partial \alpha_2} + J_{14} \frac{\partial J_{32}}{\partial \alpha_2} + J_{13} \frac{\partial J_{24}}{\partial \alpha_2} + J_{22} \frac{\partial J_{43}}{\partial \beta_1}\\ \nonumber
&+& J_{24} \frac{\partial J_{32}}{\partial \beta_1} + J_{23} \frac{\partial J_{24}}{\partial \beta_1} + J_{32} \frac{\partial J_{43}}{\partial \beta_2} + J_{34} \frac{\partial J_{32}}{\partial \beta_2} + J_{33} \frac{\partial J_{24}}{\partial \beta_2} + J_{42} \frac{\partial J_{43}}{\partial b} + J_{44} \frac{\partial J_{32}}{\partial b} + J_{43} \frac{\partial J_{24}}{\partial b} \bigg)\\ \nonumber
&=&\frac{4\left[d_e^2\left(\beta_1(t)\frac{\partial \gamma}{\partial \alpha_1} + \beta_2(t)\frac{\partial \gamma}{\partial \alpha_2}\right) + \alpha_1(t)\frac{\partial \gamma}{\partial \beta_1} + \alpha_2(t)\frac{\partial \gamma}{\partial \beta_2}\right]}{d_e^4},
\end{eqnarray}
\begin{eqnarray}
\Sigma_{301} &=& \bigg( J_{03} \frac{\partial J_{01}}{\partial \alpha_1} + J_{00} \frac{\partial J_{13}}{\partial \alpha_1} + J_{01} \frac{\partial J_{30}}{\partial \alpha_1} + J_{13} \frac{\partial J_{01}}{\partial \alpha_2} + J_{10} \frac{\partial J_{13}}{\partial \alpha_2} + J_{11} \frac{\partial J_{30}}{\partial \alpha_2} + J_{23} \frac{\partial J_{01}}{\partial \beta_1}\\ \nonumber
&+& J_{20} \frac{\partial J_{13}}{\partial \beta_1} + J_{21} \frac{\partial J_{30}}{\partial \beta_1} + J_{33} \frac{\partial J_{01}}{\partial \beta_2} + J_{30} \frac{\partial J_{13}}{\partial \beta_2} + J_{31} \frac{\partial J_{30}}{\partial \beta_2} + J_{43} \frac{\partial J_{01}}{\partial b} + J_{40} \frac{\partial J_{13}}{\partial b} + J_{41} \frac{\partial J_{30}}{\partial b} \bigg)\\ \nonumber
&=&\frac{4\left[-\left(d_i - \frac{\partial \gamma}{\partial b}\right)\alpha_2(t) + \gamma \frac{\partial \gamma}{\partial \beta_1}\right]}{d_e^2},
\end{eqnarray}
\begin{eqnarray}
\Sigma_{302} &=& \bigg( J_{03} \frac{\partial J_{02}}{\partial \alpha_1} + J_{00} \frac{\partial J_{23}}{\partial \alpha_1} + J_{02} \frac{\partial J_{30}}{\partial \alpha_1} + J_{13} \frac{\partial J_{02}}{\partial \alpha_2} + J_{10} \frac{\partial J_{23}}{\partial \alpha_2} + J_{12} \frac{\partial J_{30}}{\partial \alpha_2} + J_{23} \frac{\partial J_{02}}{\partial \beta_1}\\ \nonumber
&+& J_{20} \frac{\partial J_{23}}{\partial \beta_1} + J_{22} \frac{\partial J_{30}}{\partial \beta_1} + J_{33} \frac{\partial J_{02}}{\partial \beta_2} + J_{30} \frac{\partial J_{23}}{\partial \beta_2} + J_{32} \frac{\partial J_{30}}{\partial \beta_2} + J_{43} \frac{\partial J_{02}}{\partial b} + J_{40} \frac{\partial J_{23}}{\partial b} + J_{42} \frac{\partial J_{30}}{\partial b} \bigg)\\ \nonumber
&=&\frac{4\left[\left(d_i b(t) - \gamma\right)\frac{\partial \gamma}{\partial \alpha_2} + \beta_1(t)\frac{\partial \gamma}{\partial b}\right]}{d_e^2},
\end{eqnarray}
\begin{eqnarray}
\Sigma_{304} &=& \bigg( J_{03} \frac{\partial J_{04}}{\partial \alpha_1} + J_{00} \frac{\partial J_{43}}{\partial \alpha_1} + J_{04} \frac{\partial J_{30}}{\partial \alpha_1} + J_{13} \frac{\partial J_{04}}{\partial \alpha_2} + J_{10} \frac{\partial J_{43}}{\partial \alpha_2} + J_{14} \frac{\partial J_{30}}{\partial \alpha_2} + J_{23} \frac{\partial J_{04}}{\partial \beta_1}\\ \nonumber
&+& J_{20} \frac{\partial J_{43}}{\partial \beta_1} + J_{24} \frac{\partial J_{30}}{\partial \beta_1} + J_{33} \frac{\partial J_{04}}{\partial \beta_2} + J_{30} \frac{\partial J_{43}}{\partial \beta_2} + J_{34} \frac{\partial J_{30}}{\partial \beta_2} + J_{43} \frac{\partial J_{04}}{\partial b} + J_{40} \frac{\partial J_{43}}{\partial b} + J_{44} \frac{\partial J_{30}}{\partial b} \bigg)\\ \nonumber
&=&\frac{4\left[d_i b(t) - 2 \gamma\right]}{d_e^2},
\end{eqnarray}
\begin{eqnarray}
\Sigma_{310} &=& \bigg( J_{03} \frac{\partial J_{10}}{\partial \alpha_1} + J_{01} \frac{\partial J_{03}}{\partial \alpha_1} + J_{00} \frac{\partial J_{31}}{\partial \alpha_1} + J_{13} \frac{\partial J_{10}}{\partial \alpha_2} + J_{11} \frac{\partial J_{03}}{\partial \alpha_2} + J_{10} \frac{\partial J_{31}}{\partial \alpha_2} + J_{23} \frac{\partial J_{10}}{\partial \beta_1}\\ \nonumber
&+& J_{21} \frac{\partial J_{03}}{\partial \beta_1} + J_{20} \frac{\partial J_{31}}{\partial \beta_1} + J_{33} \frac{\partial J_{10}}{\partial \beta_2} + J_{31} \frac{\partial J_{03}}{\partial \beta_2} + J_{30} \frac{\partial J_{31}}{\partial \beta_2} + J_{43} \frac{\partial J_{10}}{\partial b} + J_{41} \frac{\partial J_{03}}{\partial b} + J_{40} \frac{\partial J_{31}}{\partial b} \bigg)\\ \nonumber
&=&\frac{4\left[\left(d_i - \frac{\partial \gamma}{\partial b}\right)\alpha_2(t) - \gamma \frac{\partial \gamma}{\partial \beta_1}\right]}{d_e^2},
\end{eqnarray}
\begin{eqnarray}
\Sigma_{312} &=& \bigg( J_{03} \frac{\partial J_{12}}{\partial \alpha_1} + J_{01} \frac{\partial J_{23}}{\partial \alpha_1} + J_{02} \frac{\partial J_{31}}{\partial \alpha_1} + J_{13} \frac{\partial J_{12}}{\partial \alpha_2} + J_{11} \frac{\partial J_{23}}{\partial \alpha_2} + J_{12} \frac{\partial J_{31}}{\partial \alpha_2} + J_{23} \frac{\partial J_{12}}{\partial \beta_1}\\ \nonumber
&+& J_{21} \frac{\partial J_{23}}{\partial \beta_1} + J_{22} \frac{\partial J_{31}}{\partial \beta_1} + J_{33} \frac{\partial J_{12}}{\partial \beta_2} + J_{31} \frac{\partial J_{23}}{\partial \beta_2} + J_{32} \frac{\partial J_{31}}{\partial \beta_2} + J_{43} \frac{\partial J_{12}}{\partial b} + J_{41} \frac{\partial J_{23}}{\partial b} + J_{42} \frac{\partial J_{31}}{\partial b} \bigg)\\ \nonumber
&=&\frac{4\left[-\left(d_i b(t) - \gamma\right)\frac{\partial \gamma}{\partial \alpha_1} + \beta_2(t)\frac{\partial \gamma}{\partial b}\right]}{d_e^2},
\end{eqnarray}
\begin{eqnarray}
\Sigma_{314} &=& \bigg( J_{03} \frac{\partial J_{14}}{\partial \alpha_1} + J_{01} \frac{\partial J_{43}}{\partial \alpha_1} + J_{04} \frac{\partial J_{31}}{\partial \alpha_1} + J_{13} \frac{\partial J_{14}}{\partial \alpha_2} + J_{11} \frac{\partial J_{43}}{\partial \alpha_2} + J_{14} \frac{\partial J_{31}}{\partial \alpha_2} + J_{23} \frac{\partial J_{14}}{\partial \beta_1}\\ \nonumber
&+& J_{21} \frac{\partial J_{43}}{\partial \beta_1} + J_{24} \frac{\partial J_{31}}{\partial \beta_1} + J_{33} \frac{\partial J_{14}}{\partial \beta_2} + J_{31} \frac{\partial J_{43}}{\partial \beta_2} + J_{34} \frac{\partial J_{31}}{\partial \beta_2} + J_{43} \frac{\partial J_{14}}{\partial b} + J_{41} \frac{\partial J_{43}}{\partial b} + J_{44} \frac{\partial J_{31}}{\partial b} \bigg)\\ \nonumber
&=&0,
\end{eqnarray}
\begin{eqnarray}
\Sigma_{320} &=& \bigg( J_{03} \frac{\partial J_{20}}{\partial \alpha_1} + J_{02} \frac{\partial J_{03}}{\partial \alpha_1} + J_{00} \frac{\partial J_{32}}{\partial \alpha_1} + J_{13} \frac{\partial J_{20}}{\partial \alpha_2} + J_{12} \frac{\partial J_{03}}{\partial \alpha_2} + J_{10} \frac{\partial J_{32}}{\partial \alpha_2} + J_{23} \frac{\partial J_{20}}{\partial \beta_1}\\ \nonumber
&+& J_{22} \frac{\partial J_{03}}{\partial \beta_1} + J_{20} \frac{\partial J_{32}}{\partial \beta_1} + J_{33} \frac{\partial J_{20}}{\partial \beta_2} + J_{32} \frac{\partial J_{03}}{\partial \beta_2} + J_{30} \frac{\partial J_{32}}{\partial \beta_2} + J_{43} \frac{\partial J_{20}}{\partial b} + J_{42} \frac{\partial J_{03}}{\partial b} + J_{40} \frac{\partial J_{32}}{\partial b} \bigg)\\ \nonumber
&=&\frac{4\left[-\left(d_i b(t) - \gamma\right)\frac{\partial \gamma}{\partial \alpha_2} - \beta_1(t)\frac{\partial \gamma}{\partial b}\right]}{d_e^2},
\end{eqnarray}
\begin{eqnarray}
\Sigma_{321} &=& \bigg( J_{03} \frac{\partial J_{21}}{\partial \alpha_1} + J_{02} \frac{\partial J_{13}}{\partial \alpha_1} + J_{01} \frac{\partial J_{32}}{\partial \alpha_1} + J_{13} \frac{\partial J_{21}}{\partial \alpha_2} + J_{12} \frac{\partial J_{13}}{\partial \alpha_2} + J_{11} \frac{\partial J_{32}}{\partial \alpha_2} + J_{23} \frac{\partial J_{21}}{\partial \beta_1}\\ \nonumber
&+& J_{22} \frac{\partial J_{13}}{\partial \beta_1} + J_{21} \frac{\partial J_{32}}{\partial \beta_1} + J_{33} \frac{\partial J_{21}}{\partial \beta_2} + J_{32} \frac{\partial J_{13}}{\partial \beta_2} + J_{31} \frac{\partial J_{32}}{\partial \beta_2} + J_{43} \frac{\partial J_{21}}{\partial b} + J_{42} \frac{\partial J_{13}}{\partial b} + J_{41} \frac{\partial J_{32}}{\partial b} \bigg)\\ \nonumber
&=&\frac{4\left[\left(d_i b(t) - \gamma\right)\frac{\partial \gamma}{\partial \alpha_1} - \beta_2(t)\frac{\partial \gamma}{\partial b}\right]}{d_e^2},
\end{eqnarray}
\begin{eqnarray}
\Sigma_{324} &=& \bigg( J_{03} \frac{\partial J_{24}}{\partial \alpha_1} + J_{02} \frac{\partial J_{43}}{\partial \alpha_1} + J_{04} \frac{\partial J_{32}}{\partial \alpha_1} + J_{13} \frac{\partial J_{24}}{\partial \alpha_2} + J_{12} \frac{\partial J_{43}}{\partial \alpha_2} + J_{14} \frac{\partial J_{32}}{\partial \alpha_2} + J_{23} \frac{\partial J_{24}}{\partial \beta_1}\\ \nonumber
&+& J_{22} \frac{\partial J_{43}}{\partial \beta_1} + J_{24} \frac{\partial J_{32}}{\partial \beta_1} + J_{33} \frac{\partial J_{24}}{\partial \beta_2} + J_{32} \frac{\partial J_{43}}{\partial \beta_2} + J_{34} \frac{\partial J_{32}}{\partial \beta_2} + J_{43} \frac{\partial J_{24}}{\partial b} + J_{42} \frac{\partial J_{43}}{\partial b} + J_{44} \frac{\partial J_{32}}{\partial b} \bigg)\\ \nonumber
&=&\frac{4\left[d_e^2\left(\beta_1(t)\frac{\partial \gamma}{\partial \alpha_1} + \beta_2(t)\frac{\partial \gamma}{\partial \alpha_2}\right) + \alpha_1(t)\frac{\partial \gamma}{\partial \beta_1} + \alpha_2(t)\frac{\partial \gamma}{\partial \beta_2}\right]}{d_e^4},
\end{eqnarray}
\begin{eqnarray}
\Sigma_{340} &=& \bigg( J_{03} \frac{\partial J_{40}}{\partial \alpha_1} + J_{04} \frac{\partial J_{03}}{\partial \alpha_1} + J_{00} \frac{\partial J_{34}}{\partial \alpha_1} + J_{13} \frac{\partial J_{40}}{\partial \alpha_2} + J_{14} \frac{\partial J_{03}}{\partial \alpha_2} + J_{10} \frac{\partial J_{34}}{\partial \alpha_2} + J_{23} \frac{\partial J_{40}}{\partial \beta_1}\\ \nonumber
&+& J_{24} \frac{\partial J_{03}}{\partial \beta_1} + J_{20} \frac{\partial J_{34}}{\partial \beta_1} + J_{33} \frac{\partial J_{40}}{\partial \beta_2} + J_{34} \frac{\partial J_{03}}{\partial \beta_2} + J_{30} \frac{\partial J_{34}}{\partial \beta_2} + J_{43} \frac{\partial J_{40}}{\partial b} + J_{44} \frac{\partial J_{03}}{\partial b} + J_{40} \frac{\partial J_{34}}{\partial b} \bigg)\\ \nonumber
&=& \frac{4\left[-d_i b(t) + 2 \gamma\right]}{d_e^2},
\end{eqnarray}
\begin{eqnarray}
\Sigma_{341} &=& \bigg( J_{03} \frac{\partial J_{41}}{\partial \alpha_1} + J_{04} \frac{\partial J_{13}}{\partial \alpha_1} + J_{01} \frac{\partial J_{34}}{\partial \alpha_1} + J_{13} \frac{\partial J_{41}}{\partial \alpha_2} + J_{14} \frac{\partial J_{13}}{\partial \alpha_2} + J_{11} \frac{\partial J_{34}}{\partial \alpha_2} + J_{23} \frac{\partial J_{41}}{\partial \beta_1}\\ \nonumber
&+& J_{24} \frac{\partial J_{13}}{\partial \beta_1} + J_{21} \frac{\partial J_{34}}{\partial \beta_1} + J_{33} \frac{\partial J_{41}}{\partial \beta_2} + J_{34} \frac{\partial J_{13}}{\partial \beta_2} + J_{31} \frac{\partial J_{34}}{\partial \beta_2} + J_{43} \frac{\partial J_{41}}{\partial b} + J_{44} \frac{\partial J_{13}}{\partial b} + J_{41} \frac{\partial J_{34}}{\partial b} \bigg)\\ \nonumber
&=&0,
\end{eqnarray}
\begin{eqnarray}
\Sigma_{342} &=& \bigg( J_{03} \frac{\partial J_{42}}{\partial \alpha_1} + J_{04} \frac{\partial J_{23}}{\partial \alpha_1} + J_{02} \frac{\partial J_{34}}{\partial \alpha_1} + J_{13} \frac{\partial J_{42}}{\partial \alpha_2} + J_{14} \frac{\partial J_{23}}{\partial \alpha_2} + J_{12} \frac{\partial J_{34}}{\partial \alpha_2} + J_{23} \frac{\partial J_{42}}{\partial \beta_1}\\ \nonumber
&+& J_{24} \frac{\partial J_{23}}{\partial \beta_1} + J_{22} \frac{\partial J_{34}}{\partial \beta_1} + J_{33} \frac{\partial J_{42}}{\partial \beta_2} + J_{34} \frac{\partial J_{23}}{\partial \beta_2} + J_{32} \frac{\partial J_{34}}{\partial \beta_2} + J_{43} \frac{\partial J_{42}}{\partial b} + J_{44} \frac{\partial J_{23}}{\partial b} + J_{42} \frac{\partial J_{34}}{\partial b} \bigg)\\ \nonumber
&=&\frac{4\left[d_e^2\left(-\beta_1(t)\frac{\partial \gamma}{\partial \alpha_1} - \beta_2(t)\frac{\partial \gamma}{\partial \alpha_2}\right) - \alpha_1(t)\frac{\partial \gamma}{\partial \beta_1} - \alpha_2(t)\frac{\partial \gamma}{\partial \beta_2}\right]}{d_e^4},
\end{eqnarray}
\begin{eqnarray}
\Sigma_{401} &=& \bigg( J_{04} \frac{\partial J_{01}}{\partial \alpha_1} + J_{00} \frac{\partial J_{14}}{\partial \alpha_1} + J_{01} \frac{\partial J_{40}}{\partial \alpha_1} + J_{14} \frac{\partial J_{01}}{\partial \alpha_2} + J_{10} \frac{\partial J_{14}}{\partial \alpha_2} + J_{11} \frac{\partial J_{40}}{\partial \alpha_2} + J_{24} \frac{\partial J_{01}}{\partial \beta_1}\\ \nonumber
&+& J_{20} \frac{\partial J_{14}}{\partial \beta_1} + J_{21} \frac{\partial J_{40}}{\partial \beta_1} + J_{34} \frac{\partial J_{01}}{\partial \beta_2} + J_{30} \frac{\partial J_{14}}{\partial \beta_2} + J_{31} \frac{\partial J_{40}}{\partial \beta_2} + J_{44} \frac{\partial J_{01}}{\partial b} + J_{40} \frac{\partial J_{14}}{\partial b} + J_{41} \frac{\partial J_{40}}{\partial b} \bigg)\\ \nonumber
&=&\frac{4\left[d_e^2\left(-\beta_1(t)\frac{\partial \gamma}{\partial \alpha_1} - \beta_2(t)\frac{\partial \gamma}{\partial \alpha_2}\right) - \alpha_1(t)\frac{\partial \gamma}{\partial \beta_1} - \alpha_2(t)\frac{\partial \gamma}{\partial \beta_2}\right]}{d_e^2},
\end{eqnarray}
\begin{eqnarray}
\Sigma_{402} &=& \bigg( J_{04} \frac{\partial J_{02}}{\partial \alpha_1} + J_{00} \frac{\partial J_{24}}{\partial \alpha_1} + J_{02} \frac{\partial J_{40}}{\partial \alpha_1} + J_{14} \frac{\partial J_{02}}{\partial \alpha_2} + J_{10} \frac{\partial J_{24}}{\partial \alpha_2} + J_{12} \frac{\partial J_{40}}{\partial \alpha_2} + J_{24} \frac{\partial J_{02}}{\partial \beta_1}\\ \nonumber
&+& J_{20} \frac{\partial J_{24}}{\partial \beta_1} + J_{22} \frac{\partial J_{40}}{\partial \beta_1} + J_{34} \frac{\partial J_{02}}{\partial \beta_2} + J_{30} \frac{\partial J_{24}}{\partial \beta_2} + J_{32} \frac{\partial J_{40}}{\partial \beta_2} + J_{44} \frac{\partial J_{02}}{\partial b} + J_{40} \frac{\partial J_{24}}{\partial b} + J_{42} \frac{\partial J_{40}}{\partial b} \bigg)\\ \nonumber
&=&0,
\end{eqnarray}
\begin{eqnarray}
\Sigma_{403} &=& \bigg( J_{04} \frac{\partial J_{03}}{\partial \alpha_1} + J_{00} \frac{\partial J_{34}}{\partial \alpha_1} + J_{03} \frac{\partial J_{40}}{\partial \alpha_1} + J_{14} \frac{\partial J_{03}}{\partial \alpha_2} + J_{10} \frac{\partial J_{34}}{\partial \alpha_2} + J_{13} \frac{\partial J_{40}}{\partial \alpha_2} + J_{24} \frac{\partial J_{03}}{\partial \beta_1}\\ \nonumber
&+& J_{20} \frac{\partial J_{34}}{\partial \beta_1} + J_{23} \frac{\partial J_{40}}{\partial \beta_1} + J_{34} \frac{\partial J_{03}}{\partial \beta_2} + J_{30} \frac{\partial J_{34}}{\partial \beta_2} + J_{33} \frac{\partial J_{40}}{\partial \beta_2} + J_{44} \frac{\partial J_{03}}{\partial b} + J_{40} \frac{\partial J_{34}}{\partial b} + J_{43} \frac{\partial J_{40}}{\partial b} \bigg)\\ \nonumber
&=&\frac{4\left[-d_i b(t) + 2 \gamma\right]}{d_e^2},
\end{eqnarray}
\begin{eqnarray}
\Sigma_{410} &=& \bigg( J_{04} \frac{\partial J_{10}}{\partial \alpha_1} + J_{01} \frac{\partial J_{04}}{\partial \alpha_1} + J_{00} \frac{\partial J_{41}}{\partial \alpha_1} + J_{14} \frac{\partial J_{10}}{\partial \alpha_2} + J_{11} \frac{\partial J_{04}}{\partial \alpha_2} + J_{10} \frac{\partial J_{41}}{\partial \alpha_2} + J_{24} \frac{\partial J_{10}}{\partial \beta_1}\\ \nonumber
&+& J_{21} \frac{\partial J_{04}}{\partial \beta_1} + J_{20} \frac{\partial J_{41}}{\partial \beta_1} + J_{34} \frac{\partial J_{10}}{\partial \beta_2} + J_{31} \frac{\partial J_{04}}{\partial \beta_2} + J_{30} \frac{\partial J_{41}}{\partial \beta_2} + J_{44} \frac{\partial J_{10}}{\partial b} + J_{41} \frac{\partial J_{04}}{\partial b} + J_{40} \frac{\partial J_{41}}{\partial b} \bigg)\\ \nonumber
&=&\frac{4\left[d_e^2\left(\beta_1(t)\frac{\partial \gamma}{\partial \alpha_1} + \beta_2(t)\frac{\partial \gamma}{\partial \alpha_2}\right) + \alpha_1(t)\frac{\partial \gamma}{\partial \beta_1} + \alpha_2(t)\frac{\partial \gamma}{\partial \beta_2}\right]}{d_e^2},
\end{eqnarray}
\begin{eqnarray}
\Sigma_{412} &=& \bigg( J_{04} \frac{\partial J_{12}}{\partial \alpha_1} + J_{01} \frac{\partial J_{24}}{\partial \alpha_1} + J_{02} \frac{\partial J_{41}}{\partial \alpha_1} + J_{14} \frac{\partial J_{12}}{\partial \alpha_2} + J_{11} \frac{\partial J_{24}}{\partial \alpha_2} + J_{12} \frac{\partial J_{41}}{\partial \alpha_2} + J_{24} \frac{\partial J_{12}}{\partial \beta_1}\\ \nonumber
&+& J_{21} \frac{\partial J_{24}}{\partial \beta_1} + J_{22} \frac{\partial J_{41}}{\partial \beta_1} + J_{34} \frac{\partial J_{12}}{\partial \beta_2} + J_{31} \frac{\partial J_{24}}{\partial \beta_2} + J_{32} \frac{\partial J_{41}}{\partial \beta_2} + J_{44} \frac{\partial J_{12}}{\partial b} + J_{41} \frac{\partial J_{24}}{\partial b} + J_{42} \frac{\partial J_{41}}{\partial b} \bigg)\\ \nonumber
&=&\frac{4\left[d_i b(t) - 2 \gamma\right]}{d_e^2},
\end{eqnarray}
\begin{eqnarray}
\Sigma_{413} &=& \bigg( J_{04} \frac{\partial J_{13}}{\partial \alpha_1} + J_{01} \frac{\partial J_{34}}{\partial \alpha_1} + J_{03} \frac{\partial J_{41}}{\partial \alpha_1} + J_{14} \frac{\partial J_{13}}{\partial \alpha_2} + J_{11} \frac{\partial J_{34}}{\partial \alpha_2} + J_{13} \frac{\partial J_{41}}{\partial \alpha_2} + J_{24} \frac{\partial J_{13}}{\partial \beta_1}\\ \nonumber
&+& J_{21} \frac{\partial J_{34}}{\partial \beta_1} + J_{23} \frac{\partial J_{41}}{\partial \beta_1} + J_{34} \frac{\partial J_{13}}{\partial \beta_2} + J_{31} \frac{\partial J_{34}}{\partial \beta_2} + J_{33} \frac{\partial J_{41}}{\partial \beta_2} + J_{44} \frac{\partial J_{13}}{\partial b} + J_{41} \frac{\partial J_{34}}{\partial b} + J_{43} \frac{\partial J_{41}}{\partial b} \bigg)\\ \nonumber
&=&0,
\end{eqnarray}
\begin{eqnarray}
\Sigma_{420} &=& \bigg( J_{04} \frac{\partial J_{20}}{\partial \alpha_1} + J_{02} \frac{\partial J_{04}}{\partial \alpha_1} + J_{00} \frac{\partial J_{42}}{\partial \alpha_1} + J_{14} \frac{\partial J_{20}}{\partial \alpha_2} + J_{12} \frac{\partial J_{04}}{\partial \alpha_2} + J_{10} \frac{\partial J_{42}}{\partial \alpha_2} + J_{24} \frac{\partial J_{20}}{\partial \beta_1}\\ \nonumber
&+& J_{22} \frac{\partial J_{04}}{\partial \beta_1} + J_{20} \frac{\partial J_{42}}{\partial \beta_1} + J_{34} \frac{\partial J_{20}}{\partial \beta_2} + J_{32} \frac{\partial J_{04}}{\partial \beta_2} + J_{30} \frac{\partial J_{42}}{\partial \beta_2} + J_{44} \frac{\partial J_{20}}{\partial b} + J_{42} \frac{\partial J_{04}}{\partial b} + J_{40} \frac{\partial J_{42}}{\partial b} \bigg)\\ \nonumber
&=&0,
\end{eqnarray}
\begin{eqnarray}
\Sigma_{421} &=& \bigg( J_{04} \frac{\partial J_{21}}{\partial \alpha_1} + J_{02} \frac{\partial J_{14}}{\partial \alpha_1} + J_{01} \frac{\partial J_{42}}{\partial \alpha_1} + J_{14} \frac{\partial J_{21}}{\partial \alpha_2} + J_{12} \frac{\partial J_{14}}{\partial \alpha_2} + J_{11} \frac{\partial J_{42}}{\partial \alpha_2} + J_{24} \frac{\partial J_{21}}{\partial \beta_1}\\ \nonumber
&+& J_{22} \frac{\partial J_{14}}{\partial \beta_1} + J_{21} \frac{\partial J_{42}}{\partial \beta_1} + J_{34} \frac{\partial J_{21}}{\partial \beta_2} + J_{32} \frac{\partial J_{14}}{\partial \beta_2} + J_{31} \frac{\partial J_{42}}{\partial \beta_2} + J_{44} \frac{\partial J_{21}}{\partial b} + J_{42} \frac{\partial J_{14}}{\partial b} + J_{41} \frac{\partial J_{42}}{\partial b} \bigg)\\ \nonumber
&=&\frac{4\left[-d_i b(t) + 2 \gamma\right]}{d_e^2},
\end{eqnarray}
\begin{eqnarray}
\Sigma_{423} &=& \bigg( J_{04} \frac{\partial J_{23}}{\partial \alpha_1} + J_{02} \frac{\partial J_{34}}{\partial \alpha_1} + J_{03} \frac{\partial J_{42}}{\partial \alpha_1} + J_{14} \frac{\partial J_{23}}{\partial \alpha_2} + J_{12} \frac{\partial J_{34}}{\partial \alpha_2} + J_{13} \frac{\partial J_{42}}{\partial \alpha_2} + J_{24} \frac{\partial J_{23}}{\partial \beta_1}\\ \nonumber
&+& J_{22} \frac{\partial J_{34}}{\partial \beta_1} + J_{23} \frac{\partial J_{42}}{\partial \beta_1} + J_{34} \frac{\partial J_{23}}{\partial \beta_2} + J_{32} \frac{\partial J_{34}}{\partial \beta_2} + J_{33} \frac{\partial J_{42}}{\partial \beta_2} + J_{44} \frac{\partial J_{23}}{\partial b} + J_{42} \frac{\partial J_{34}}{\partial b} + J_{43} \frac{\partial J_{42}}{\partial b} \bigg)\\ \nonumber
&=&\frac{4\left[d_e^2\left(-\beta_1(t)\frac{\partial \gamma}{\partial \alpha_1} - \beta_2(t)\frac{\partial \gamma}{\partial \alpha_2}\right) - \alpha_1(t)\frac{\partial \gamma}{\partial \beta_1} - \alpha_2(t)\frac{\partial \gamma}{\partial \beta_2}\right]}{d_e^4},
\end{eqnarray}
\begin{eqnarray}
\Sigma_{430} &=& \bigg( J_{04} \frac{\partial J_{30}}{\partial \alpha_1} + J_{03} \frac{\partial J_{04}}{\partial \alpha_1} + J_{00} \frac{\partial J_{43}}{\partial \alpha_1} + J_{14} \frac{\partial J_{30}}{\partial \alpha_2} + J_{13} \frac{\partial J_{04}}{\partial \alpha_2} + J_{10} \frac{\partial J_{43}}{\partial \alpha_2} + J_{24} \frac{\partial J_{30}}{\partial \beta_1}\\ \nonumber
&+& J_{23} \frac{\partial J_{04}}{\partial \beta_1} + J_{20} \frac{\partial J_{43}}{\partial \beta_1} + J_{34} \frac{\partial J_{30}}{\partial \beta_2} + J_{33} \frac{\partial J_{04}}{\partial \beta_2} + J_{30} \frac{\partial J_{43}}{\partial \beta_2} + J_{44} \frac{\partial J_{30}}{\partial b} + J_{43} \frac{\partial J_{04}}{\partial b} + J_{40} \frac{\partial J_{43}}{\partial b} \bigg)\\ \nonumber
&=&\frac{4\left[d_i b(t) - 2 \gamma\right]}{d_e^2},
\end{eqnarray}
\begin{eqnarray}
\Sigma_{431} &=& \bigg( J_{04} \frac{\partial J_{31}}{\partial \alpha_1} + J_{03} \frac{\partial J_{14}}{\partial \alpha_1} + J_{01} \frac{\partial J_{43}}{\partial \alpha_1} + J_{14} \frac{\partial J_{31}}{\partial \alpha_2} + J_{13} \frac{\partial J_{14}}{\partial \alpha_2} + J_{11} \frac{\partial J_{43}}{\partial \alpha_2} + J_{24} \frac{\partial J_{31}}{\partial \beta_1}\\ \nonumber
&+& J_{23} \frac{\partial J_{14}}{\partial \beta_1} + J_{21} \frac{\partial J_{43}}{\partial \beta_1} + J_{34} \frac{\partial J_{31}}{\partial \beta_2} + J_{33} \frac{\partial J_{14}}{\partial \beta_2} + J_{31} \frac{\partial J_{43}}{\partial \beta_2} + J_{44} \frac{\partial J_{31}}{\partial b} + J_{43} \frac{\partial J_{14}}{\partial b} + J_{41} \frac{\partial J_{43}}{\partial b} \bigg)\\ \nonumber
&=&0,
\end{eqnarray}
\begin{eqnarray}
\Sigma_{432} &=& \bigg( J_{04} \frac{\partial J_{32}}{\partial \alpha_1} + J_{03} \frac{\partial J_{24}}{\partial \alpha_1} + J_{02} \frac{\partial J_{43}}{\partial \alpha_1} + J_{14} \frac{\partial J_{32}}{\partial \alpha_2} + J_{13} \frac{\partial J_{24}}{\partial \alpha_2} + J_{12} \frac{\partial J_{43}}{\partial \alpha_2} + J_{24} \frac{\partial J_{32}}{\partial \beta_1}\\ \nonumber
&+& J_{23} \frac{\partial J_{24}}{\partial \beta_1} + J_{22} \frac{\partial J_{43}}{\partial \beta_1} + J_{34} \frac{\partial J_{32}}{\partial \beta_2} + J_{33} \frac{\partial J_{24}}{\partial \beta_2} + J_{32} \frac{\partial J_{43}}{\partial \beta_2} + J_{44} \frac{\partial J_{32}}{\partial b} + J_{43} \frac{\partial J_{24}}{\partial b} + J_{42} \frac{\partial J_{43}}{\partial b} \bigg)\\ \nonumber
&=&\frac{4\left[d_e^2\left(\beta_1(t)\frac{\partial \gamma}{\partial \alpha_1} + \beta_2(t)\frac{\partial \gamma}{\partial \alpha_2}\right) + \alpha_1(t)\frac{\partial \gamma}{\partial \beta_1} + \alpha_2(t)\frac{\partial \gamma}{\partial \beta_2}\right]}{d_e^4},
\end{eqnarray}

On employing all these terms suitably in Section \ref{SSecJacobi}, it can be shown that the Jacobi identity is valid for the noncanonical Poisson bracket of our system.

\end{appendices}

\bibliography{Xpointcollapse}

\end{document}